\documentclass[twocolumn]{aastex62}

\usepackage{apjfonts}

\shorttitle{Nucleus of Abell 57}
\shortauthors{Bond et al.}

\newcommand{\Teff}{T_{\rm eff}}
\newcommand{\logg}{\log g}

\newcommand{\msun}{M_\sun}

\def\ariv{\ion{Ar}{4}}
\def\baii{\ion{Ba}{2}}

\def\caii{\ion{Ca}{2}}
\def\hei{\ion{He}{1}}
\def\heii{\ion{He}{2}}

\def\nii{\ion{N}{2}}

\def\neiii{\ion{Ne}{3}}

\def\oi{\ion{O}{1}}
\def\oii{\ion{O}{2}}
\def\oiii{\ion{O}{3}}

\def\sii{\ion{S}{2}}

\def\Gaia{{\it Gaia}}

\def\WISE{{\it WISE}}

\newcommand{\HST}{{\it HST}}

\newcommand{\JWST}{{\it JWST}}

\newcommand{\Swift}{{\it Swift}}
\newcommand{\cloudy}{C{\footnotesize LOUDY}}

\newcommand{\uvwone}{{\it uvw1}}
\newcommand{\uvwtwo}{{\it uvw2}}

\begin{document}

\title{Spectroscopic Survey of Faint Planetary-Nebula Nuclei. V\null. The EGB~6-Type Central Star of Abell 57\footnote{Based on observations obtained with the Hobby-Eberly Telescope (HET), which is a joint project of the University of Texas at Austin, the Pennsylvania State University, Ludwig-Maximillians-Universit\"at M\"unchen, and Georg-August Universit\"at G\"ottingen. The HET is named in honor of its principal benefactors, William P. Hobby and Robert E. Eberly.} }

\author[0000-0003-1377-7145]{Howard E. Bond}
\affil{Department of Astronomy \& Astrophysics, Pennsylvania State University, University Park, PA 16802, USA}
\affil{Space Telescope Science Institute, 
3700 San Martin Dr.,
Baltimore, MD 21218, USA}

\author[0009-0008-0451-4435]{Akshat S. Chaturvedi}
\affil{Department of Astronomy \& Astrophysics, Pennsylvania State University, University Park, PA 16802, USA}

\author[0000-0002-1328-0211]{Robin Ciardullo}
\affil{Department of Astronomy \& Astrophysics,
Pennsylvania State University, University Park, PA 16802, USA}
\affil{Institute for Gravitation and the Cosmos, The Pennsylvania
State University, University Park, PA 16802, USA}

\author[0000-0002-6428-2276]{Klaus Werner}
\affil{Institut f\"ur Astronomie und Astrophysik, Kepler Center for
  Astro and Particle Physics, Eberhard Karls Universit\"at, Sand~1, 72076
  T\"ubingen, Germany}

\author[0000-0003-2307-0629]{Gregory R. Zeimann}
\affil{Hobby-Eberly Telescope, University of Texas at Austin, Austin, TX 78712, USA}

\author[0000-0003-1817-3009]{Michael H. Siegel}
\affil{Department of Astronomy \& Astrophysics, Pennsylvania State University, University Park, PA 16802, USA}

\correspondingauthor{Howard E. Bond}
\email{heb11@psu.edu}

\begin{abstract}

During our spectroscopic survey of central stars of faint planetary nebulae (PNe), we found that the nucleus of Abell~57 exhibits strong nebular emission lines. Using synthetic narrow-band images, we show that the emission arises from an unresolved compact emission knot (CEK) coinciding with the hot (90,000~K) central star. Thus Abell~57 belongs to the rare class of ``EGB~6-type'' PNe, characterized by dense emission cores. Photometric data show that the nucleus exhibits a near-IR excess, due to a dusty companion body with the luminosity of an M0 dwarf but a temperature of $\sim$1800~K\null. Emission-line analysis reveals that the CEK is remarkably dense (electron density $\sim$$1.6\times10^7\rm\,cm^{-3}$), and has a radius of only $\sim$4.5~AU\null. The CEK suffers considerably more reddening than the central star, which itself is more reddened than the surrounding PN\null. These puzzles may suggest an interaction between the knot and central star; however, {\it Hubble Space Telescope\/} imaging of EGB~6 itself shows that its CEK lies more than $\sim$125~AU from the PN nucleus. We discuss a scenario in which a portion of the AGB wind that created the PN was captured into a dust cloud around a distant stellar companion; this cloud has survived to the present epoch, and has an atmosphere photoionized by radiation from the hot central star. However, in this picture EGB~6-type nuclei should be relatively common, yet they are actually extremely
rare; thus they may arise from a different transitory phenomenon. We suggest future observations of Abell~57 that may help unravel its mysteries.

\null\vskip 0.2in

\end{abstract}



\section{Introduction}

This is the fifth in a series of papers presenting results from a spectroscopic survey of central stars of faint Galactic planetary nebulae (PNe). The survey is carried out with the second-generation ``blue'' Low-Resolution Spectrograph (LRS2-B; \citealt{Chonis2016}) of the 10-m Hobby-Eberly Telescope (HET; \citealt{Ramsey1998,Hill2021}), located at McDonald Observatory in west Texas, USA\null. An overview of the survey, a description of the instrumentation and data-reduction procedures, target selection, and some initial results were presented in our first paper \citep[][hereafter Paper~I]{Bond2023a}. Paper~II \citep{Bond2023b} discussed the central star of the ``PN mimic'' Fr~2-30, and Paper~III \citep{Werner2024} presented three new extremely hot hydrogen-deficient planetary-nebula nuclei (PNNi). Paper~IV \citep{Bond_Pa27_2024} describes the late-type spotted and rotationally variable central star of Pa~27. About 50 PNNi have been observed to date, and a future paper will present results on a group of nuclei with fairly normal hydrogen-rich spectra. In the present fifth paper we analyze spectroscopic and photometric observations of the central star and nebula of Abell~57, which we show to be a member of the rare and poorly understood class of ``EGB\,6-type'' PNNi.

\section{EGB 6-Type PN Nuclei \label{sec:EGB6} }

The EGB~6 group of PNNi is described in a review article by \citet{FrewParker2010} as central stars that are associated with compact and unresolved (from the ground) high-density emission-line nebulae. As the authors note, about half a dozen PNNi are known or candidate members of the EGB~6 category. Several of them, including some new examples, were discussed by \citet{Miszalski2013}, who called the objects PNe with high-density cores, or HDC PNe.

The prototype of the class, EGB~6 (PN G221.5+46.3), was investigated in detail by \citet[][hereafter L13]{Liebert2013} and \citet[][hereafter B16]{BondEGB62016}, and we summarize its properties here. EGB~6 is a large (minor and major axes $11'\times13'$), ancient, and very low-surface-brightness PN, serendipitously discovered in 1978 by H.E.B. during examination of prints from the Palomar Observatory Sky Survey (POSS). The object was included in a list of faint nebulae found during subsequent systematic searches of POSS prints by \citet{Ellis1984}, who noted that the POSS photographs show a 16th-mag blue star near the center of the nebula. The recent \Gaia\/ Data Release~3\footnote{\url{https://vizier.cds.unistra.fr/viz-bin/VizieR-3?-source=I/355/gaiadr3}} (DR3; \citealt{Gaia2016, Gaia2023}) gives a parallax of $1.33\pm	0.15$~mas for this central star, implying a distance of 750~pc, and a nebular size of $2.4\times2.8$~pc (minor and major axes). The absolute magnitude of the PNN is $M_G=+6.6$.

Follow-up spectroscopy of the PN's nucleus by \citet{Ellis1984} revealed [\oiii] emission lines that are far stronger that those of the surrounding nebula; the authors demonstrated this by obtaining spectra $\sim$$10''$ away from the star, showing much fainter emission. 
\citet{Liebert1989} classified the exciting star as a hot white dwarf (WD) with Balmer and \heii\ absorption lines in its spectrum. Later \citet{Gianninas2010} found metallic absorption lines in the far-ultraviolet, leading to a spectral classification of DAOZ\null. More recently, a detailed analysis of the star's optical and ultraviolet spectra by \citet{WernerEGB62018} found an effective temperature of $\Teff=105,000\pm5,000$~K, a surface gravity of $\log g=7.4\pm0.4$, and approximately solar abundances of He and heavy elements.

\citet{Liebert1989} verified that the nebular lines in the spectrum of the EGB~6 central star arise from a compact emission knot (CEK), which is unresolved and coincides with the PNN in ground-based images. Moreover, the electron density in the CEK is remarkably high, about $2.2\times10^6 \,\rm cm^{-3}$, according to an emission-line analysis by \citet{Dopita1989}. 

In addition to the hot WD and associated CEK, there is a compact near-infrared (NIR) source located near the EGB 6 nucleus. This object was first revealed through NIR photometry by \citet{Zuckerman1991} and its presence was later confirmed by \citet{Fulbright1993} and a host of other observations (see B16 for details and references). Moreover, the source displays excess mid-infrared (MIR) flux in the {\it Spitzer Space Telescope\/} IRAC and MIPS passbands \citep{Chu2011}. 

The spectral-energy distribution (SED) of the EGB 6 PNN was assembled by B16. These authors found that the SED can be represented by a combination of four blackbodies: the hot WD, a NIR stellar source with an effective temperature of $\sim$1850~K, and two MIR dust components with temperatures of 385 and 175~K\null. The NIR source has the luminosity of a M3.5~V dwarf, but its NIR spectrum is continuous, suggesting that it is an enshrouded late-type star (or conceivably a second WD). 

High-spatial-resolution broadband and emission-line imaging of EGB~6 with the {\it Hubble Space Telescope\/} (\HST) provided another startling result: the CEK is not centered on the hot WD, but is a separate point source lying $0\farcs16$ away---a projected separation of 125~AU\null. Remarkably, the location of the CEK coincides with that of the NIR companion. These findings led B16 to suggest a scenario in which the large PN was ejected from the hot WD, and a portion of this wind was captured into a dusty optically thick accretion disk around an M dwarf companion, which survives to the present epoch. In this picture, the accretion disk is surrounded by an atmosphere of gas that is photoionized by ultraviolet radiation from the WD, producing the CEK. 

As B16 point out, a number of astrophysical puzzles are posed by the diverse phenomena seen in the EGB~6 system. It would be useful to have further examples and detailed studies of similar objects.

\section{Abell 57 \label{sec:abell57} }

Among the $\sim$50 PNNi observed to date in our HET survey, the central star of the faint nebula Abell~57 (PN~G058.6+06.1) stood out as having a spectrum with strong nebular emission lines. As we show below, this emission is unresolved and spatially coincident with the central star (in ground-based images), confirming Abell~57 as a member of the EGB~6 class.\footnote{In a conference poster, \citet{Miszalski2011} presented spectra showing that Abell~57 belongs to the EGB~6/HDC-PNe category. To our knowledge no further details were published, apart from a brief mention of Abell~57 in \citet{Miszalski2013}.} We assume hereafter that the emission lines in Abell~57 arise from a CEK, as they do in EGB~6 itself.

The nebula was discovered in the classical search of the POSS for faint PNe by \citet{Abell1966}, who noted a 17th-mag blue central star. Table~\ref{tab:DR3data} lists this star's equatorial and Galactic coordinates, parallax and proper motion, $G$ magnitude, and $G_{\rm BP}-G_{\rm RP}$ color, all taken from \Gaia\/ Data Release~3\footnote{\url{https://vizier.cds.unistra.fr/viz-bin/VizieR-3?-source=I/355/gaiadr3}. The \Gaia\/ catalog shows that a 20.0-mag red star lying $4\farcs3$ away from the central star, and a 20.6-mag star $14\farcs6$ away, have similar proper motions and may be common-proper-motion (CPM) companions. Curiously, EGB~6 also has a red CPM companion, at a separation of $27\farcs2$.} (DR3; \citealt{Gaia2016, Gaia2023}). Several images of Abell 57 obtained by amateurs are available at \url{www.astrobin.com}, including deep frames by Gary Imm\footnote{\url{https://www.astrobin.com/cl0hyf}} and Jerry Macon.\footnote{\url{https://www.astrobin.com/sk2m28}} These images show a faint elliptical nebula with diffuse edges and minor and major axes of approximately $30''\times38''$, enclosing an equatorial torus around the central star, which is brightest on the northeast, east, and southeast sides of the center.

The \Gaia\/ parallax implies a distance of $\sim$2100~pc.\footnote{A Bayesian analysis of {\it Gaia\/} EDR3 data by \citet{BailerJones2021} yields a distance of $2091^{+359}_{-271}$~pc.} At this distance, the physical minor and major axes of the faint elliptical shell are about $0.3\times0.4$~pc, suggesting that the nebula is considerably younger than the very large nebula surrounding EGB~6.

\begin{deluxetable}{lc}
\tablecaption{\Gaia\/ DR3 Data for Central Star of Abell 57 \label{tab:DR3data} }
\tablehead{
\colhead{Parameter}
&\colhead{Value}
}
\decimals
\startdata
RA (J2000) & 19 17 05.656 \\
Dec (J2000) & +25 37 32.89 \\
$l$ [deg] &  58.61 \\
$b$  [deg] &  $+6.17$ \\
Parallax [mas] & $0.470\pm0.074$ \\
$\mu_\alpha$ [mas\,yr$^{-1}$] & $-1.569\pm0.054$ \\
$\mu_\delta$ [mas\,yr$^{-1}$] & $-6.550\pm0.077$ \\
$G$ [mag] &  17.54 \\
$G_{\rm BP}-G_{\rm RP}$ [mag] & $0.42$ \\
\enddata
\end{deluxetable}

\section{Observations and Data Reduction}

\subsection{LRS2-B Spectroscopy}
\label{sec:LRS2_spectroscopy}

There are actually two LRS2 instruments, one used to obtain spectra at ``blue'' wavelengths (LRS2-B), and one that obtains ``red'' spectra (LRS2-R). We observed Abell~57 with the LRS2-B integral-field-unit (IFU) spectrograph on the HET\null. This instrument uses 280 $0\farcs6$-diameter lenslet-coupled fibers covering a $12''\times 6''$ field of view (FOV), which feed two spectrograph units via a dichroic beamsplitter.  The ``UV'' channel of LRS2-B covers the wavelength range 3640--4645~\AA\ at resolving power 1910, while the ``Orange'' channel covers 4635--6950~\AA\ at resolving power 1140. Full details of the instrument are given by \citet{Chonis2016}, and in Paper~I\null. An observation log for our LRS2-B exposures on Abell~57 is presented in Table~\ref{tab:exposures}.


\begin{deluxetable}{lc}[h]
\tablecaption{Log of HET LRS2-B Observations of Abell 57 \label{tab:exposures} }
\tablehead{
\colhead{Date}
&\colhead{Exposure}\\
\colhead{[YYYY-MM-DD]}
&\colhead{[s]}}
\decimals
\startdata
2020-11-10 & 375 \\
2021-11-08 & $2\times600$ \\
2022-07-17 & $2\times600$ \\
2023-07-21 & $2\times750$ \\
\enddata
\end{deluxetable}

As described in Paper~I, the raw LRS2-B data are first processed with the \texttt{Panacea} pipeline,\footnote{\url{https://github.com/grzeimann/Panacea}} which performs a basic bias and flat-field correction, fiber extraction, and wavelength calibration. An absolute-flux calibration is then derived from the instrument's default response curves and measures of the primary mirror's illumination and the exposure throughput, as determined from guide-camera images. \texttt{Panacea} also performs an initial median sky subtraction, source modeling, and one-dimensional spectral extraction.  However, in the special case of Abell~57, the central star is embedded in surrounding nebular emission, necessitating more complex modeling of the background.  We therefore applied \texttt{LRS2Multi}\footnote{\url{https://github.com/grzeimann/LRS2Multi}} to the un-sky-subtracted, flux-calibrated fiber spectra to perform nebular background and sky subtraction.  We masked out a 2$\arcsec$-radius region around the central star, and fit the remaining fibers with a 5th-order two-dimensional polynomial at each wavelength to model the sum of the spatially uniform sky and spatially variable nebula.  We modeled each wavelength separately, due to subtleties in the kinematics of the nebula and small variations in spectral point-spread functions (PSFs) as a function of wavelength.  At wavelengths with no strong nebular emission, the polynomial was often a near-constant value, as expected for sky emission over a small FOV\null.  After subtracting the background model, we created a synthetic continuum image for each exposure by collapsing the fiber spectra between $5100\pm50$\,\AA\null.  We then modeled the star using a two-dimensional Gaussian, and used the resulting fit to perform an optimally weighted extraction at each wavelength \citep{Horne1986}.  Finally, we normalized the extracted spectra at $5100\pm50$~\AA, with factors between 0.81 and 1.21.  These residual normalizations are due to imperfect estimates of the mirror illumination and throughput from the guider cameras.  We combined the seven individual exposures using the inverse variance weights from the error arrays to create our final one-dimensional spectrum of the central star.

We also extracted a spectrum of the surrounding nebula. For this purpose, we chose the same position in each IFU exposure---a location with a radius of $2''$, lying 4$\arcsec$ southeast of the central star---where the surface brightness of the PN in [\oiii] 5007~\AA\ is highest in the LRS2-B FOV\null. 
In this case the spectral extraction is complicated by the fact that the Abell~57 nebula has a size of $30''\times38''$, and hence there is no true region of pure sky background within the LRS2 FOV in our observations. Thus our usual sky-subtraction processing would result in some self-subtraction, which could lead to incorrect line ratios if the spectrum is spatially variable (which is known to be the case in many PNe). We therefore decided to work with an un-sky-subtracted spectrum.
We normalized the spectrum of each fiber within the $2''$-radius PN aperture at $5007 \pm 5$~\AA, with factors ranging from 0.87 to 1.12, and, similarly to the central-star extraction, we combined the individual exposures using inverse variance weights to create a final one-dimensional un-subtracted spectrum of the surrounding PN\null.

\begin{figure*}
\centering
\includegraphics[width=5in]{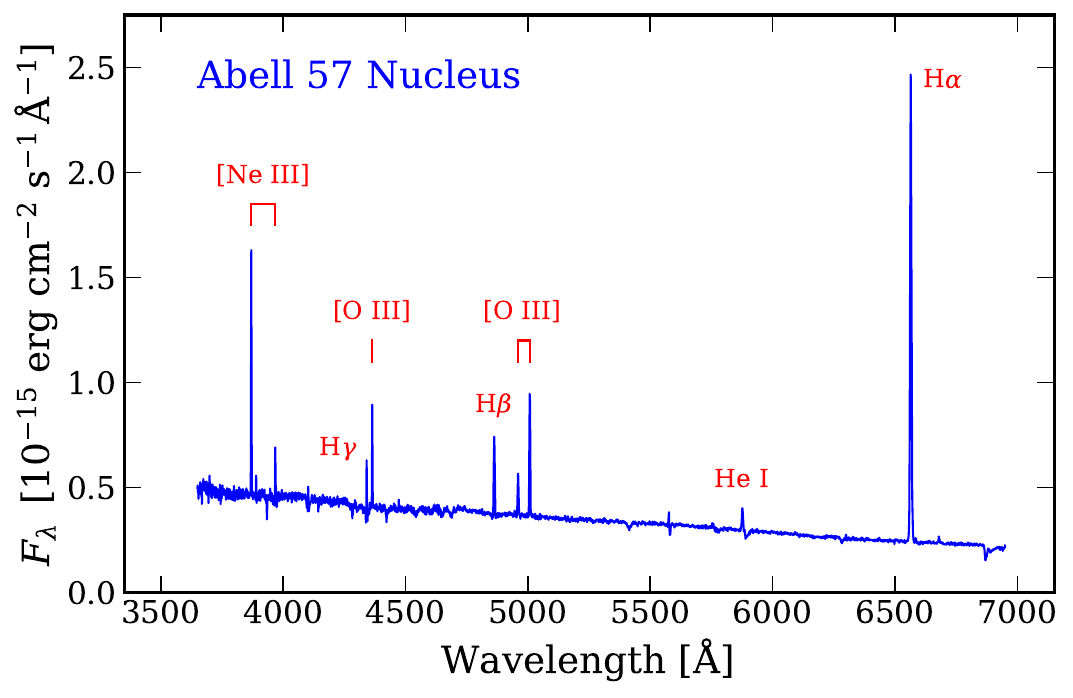}
\hbox{\hskip0.04in\includegraphics[width=4.95in]{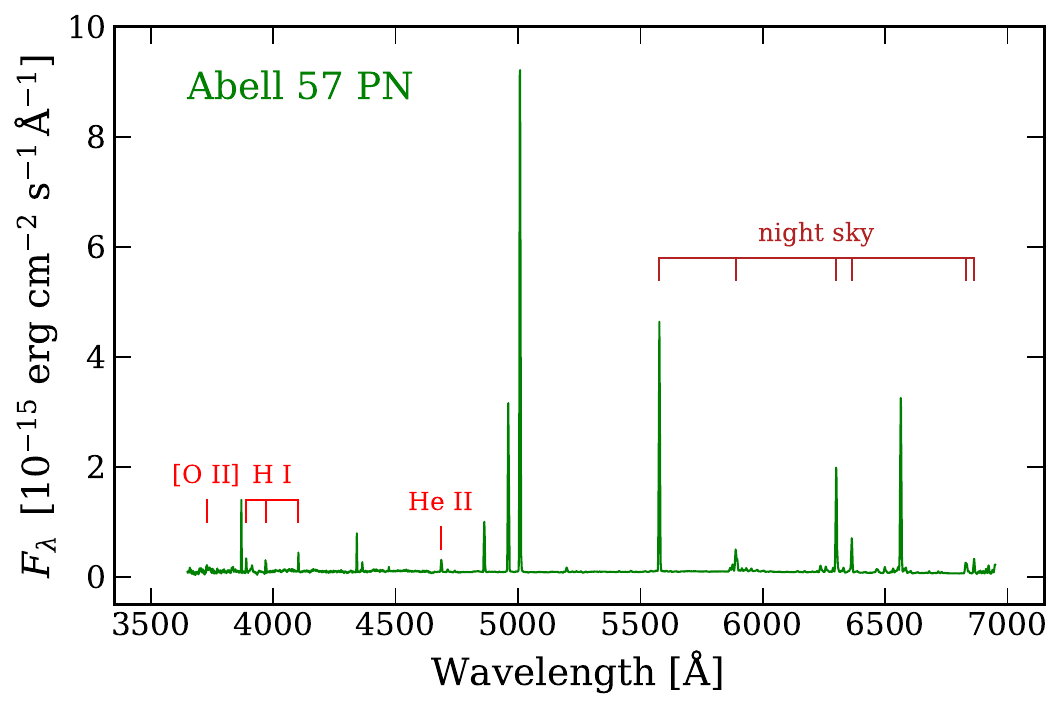}}
\caption{
Hobby-Eberly Telescope LRS2-B spectra of Abell 57's central star and compact emission-line knot (top panel), and of a bright $4''$-diameter region of the surrounding planetary nebula (bottom panel). As discussed in the text, the spectrum of the nucleus is sky- and background-subtracted, but that of the nebula is not. The stellar continuum of the hot central star is clearly seen in the top panel, with superposed emission lines from the compact knot.  Three striking differences between the spectra are that the spectrum of the nucleus shows a high flux of [\ion{O}{3}] $\lambda 4363$ relative to [\ion{O}{3}] $\lambda 5007$, the [\oiii] $\lambda\lambda$4959--5007 lines are weak relative to H$\alpha$, and the Balmer decrement is large.   As discussed in Section~\ref{sec:nebular_analysis}, the [\oiii] ratio and the quenching of [\oiii] $\lambda\lambda$4959--5007 relative to H$\alpha$ imply a very high electron density for the knot ($n_e \gtrsim 10^7$~cm$^{-3}$), and its large Balmer decrement indicates higher dust reddening for the knot than for the nebula. 
\label{fig:a57_spectra}
}
\end{figure*}

Figure~\ref{fig:a57_spectra} plots LRS2-B spectra of the central star and its CEK (top panel) and of the surrounding PN (bottom panel, showing the un-subtracted nebular spectrum).  Several striking differences between the emission-line spectra are immediately apparent, as summarized in the figure caption. These primarily reflect large differences in the extinction and nebular electron density between the CEK and the extended PN, which are discussed in Section~\ref{sec:nebular_analysis}.

\subsection{Synthetic Narrow-Band Images \label{sec:narrowbandimages} }

Since LRS2-B is an IFU spectrograph, its resulting data cube can be sliced to create synthetic spatial images, for any spectral bandpass. We used this capability to construct synthetic narrow-band (NB) emission-line-minus-continuum ``difference'' images from our LRS2-B observations, in order to test whether the emission lines in the spectrum of the Abell~57 nucleus truly arise from a compact object associated with the central star. If they do, NB images in the emission lines will show a source coinciding with, and having a PSF similar to, the central star in images taken in the adjacent continuum.

\begin{figure*}
\centering
\includegraphics[width=5in]{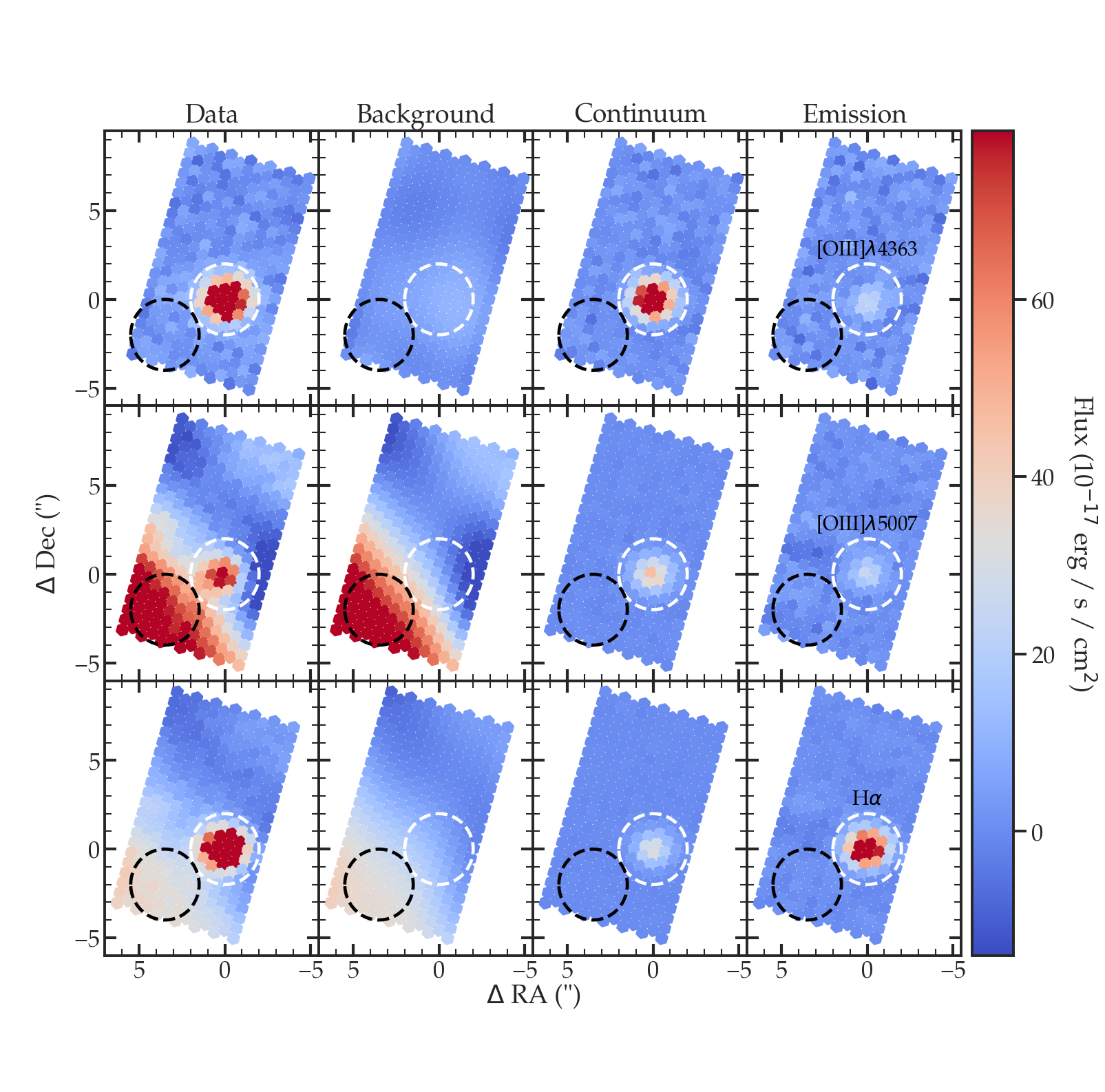}
\vskip-0.4in
\caption{
Synthetic direct images of the Abell~57 nucleus and surroundings, created from our 2022 July~17 LRS2-B IFU data for the emission lines of [\oiii] 4363~\AA\ (top row), [\oiii] 5007~\AA\ (middle row), and H$\alpha$ (bottom row).  The left-hand ``Data'' column shows the synthetic narrow-band images without background (sky and nebula) subtraction.  The ``Background'' column illustrates the models for the backgrounds in each emission line, with the star removed. The ``Continuum'' column shows images in the stellar continuum neighboring each emission line.  Finally the ``Emission'' column presents the background-subtracted image minus the continuum image.  The figure confirms that an unresolved emission-line knot coincides with the hot central star. See text for further details.  In each panel, we subtracted the 25th-percentile fiber value from the image, in order to put each of the panels on a common color range, shown on the right. The dashed white circles show the $2''$-radius extraction aperture used for the central-star spectrum plotted in the top panel of Figure~\ref{fig:a57_spectra}; the dashed black circles denote the extraction aperture used to obtain the nebular spectrum shown in the bottom panel of Figure~\ref{fig:a57_spectra}.
\label{fig:direct_images}
}
\end{figure*}

We constructed NB images for the emission lines of [\oiii] 4363~\AA, [\oiii] 5007~\AA, and H$\alpha$. The creation process is illustrated in Figure~\ref{fig:direct_images}, which shows a row for each wavelength, and four columns of synthetic images.  In each frame, the $2''$ extraction radius for the spectrum of the central star (top panel of Figure~\ref{fig:a57_spectra}) is shown as a dashed white circle, and the $2''$ extraction radius for the PN spectrum (bottom panel of Figure~\ref{fig:a57_spectra}) as a dashed black circle. 

The first column of Figure~\ref{fig:direct_images} displays the synthetic NB images without background (sky or PN) subtraction, each one using a spectral window 16~\AA\ wide. The second column presents the background models in the specified NBs, created as described in the previous subsection.  The third column shows  images in the neighboring continuum for each of the emission lines; these are averages of two images created using 16~\AA\ wide bandpasses on either side of the emission line.    The wavelength offsets for these continuum images were chosen to avoid contamination from other spectral features, and were $\pm$45~\AA\ for [\oiii] 4363~\AA, $\pm$30~\AA\ for [\oiii] 5007~\AA, and $\pm$50~\AA\ for H$\alpha$.  The final column in Figure~\ref{fig:direct_images} shows the background-subtracted emission-line images minus the continuum images. At all three wavelengths there is line emission that is co-spatial with the central star, and is much brighter than the immediately surrounding PN background. Thus these images confirm that Abell~57 is an EGB~6-type object.


In our sharpest synthetic images, obtained on 2021 November~8, the centroid locations of the CEK emission lines and of the stellar continuum coincide to within $0\farcs05$ ($\sim$100~AU at the distance of the PN) in [\oiii] 5007~\AA\ and H$\alpha$, and $0\farcs15$ in [\oiii] 4363~\AA\null. The full width at half maximum (FWHM) of the central star's image is $1\farcs55$, and the FWHMs of the emission lines agree with this measurement to within $0\farcs05$. 


\section{The Central Star of Abell 57}

\subsection{Atmospheric Analysis \label{sec:atmosph_analysis} }

We carried out an atmospheric analysis to determine properties of Abell 57's central star. It was classified as having spectral type O(H) by \cite{DeMarco2013}, based on data presented by \cite{Miszalski2011}. Our spectrum (top panel of Figure~\ref{fig:a57_spectra}, and shown in detail below) confirms this classification, as the only detected absorption lines are due to hydrogen and \heii. 

\begin{figure*}
\centering
\includegraphics[width=\textwidth]{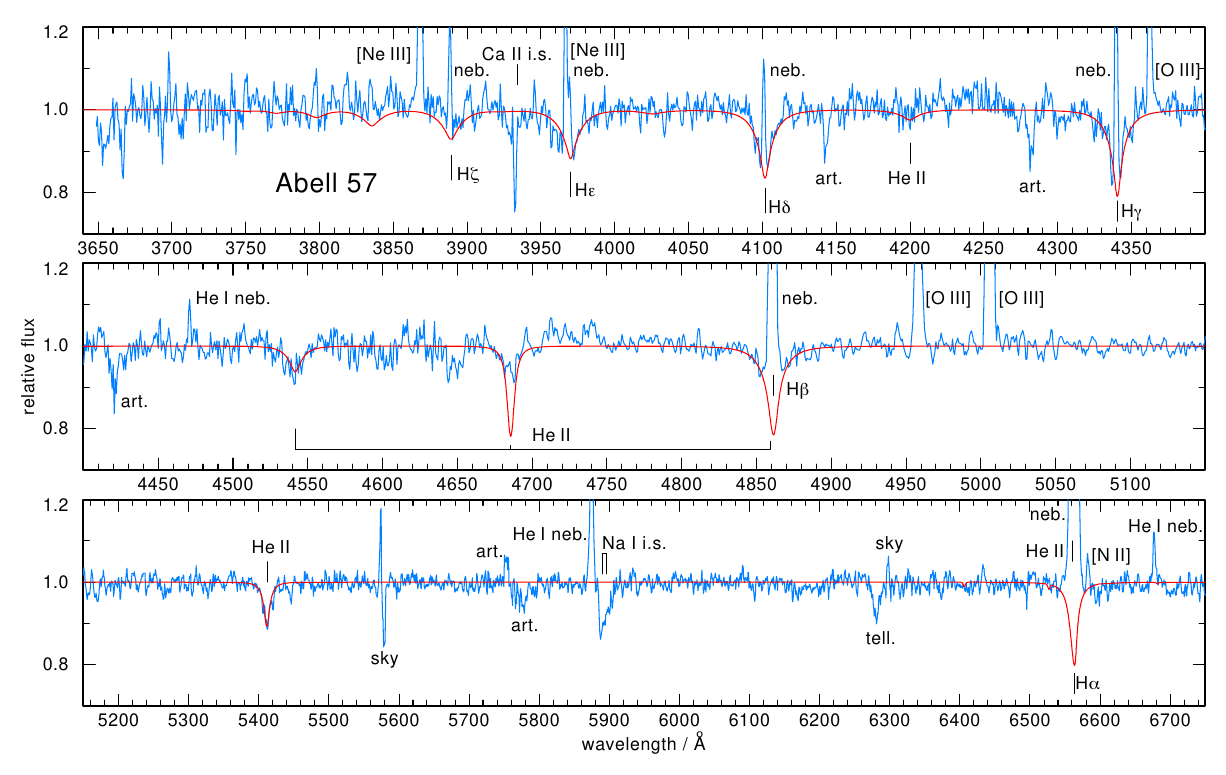}
\caption{
Normalized observed spectrum of the Abell~57 central star (blue line), superposed with our final theoretical model (red line) with $\Teff$ = 90\,000\,K, $\logg$ = 6.0, and $n_{\rm He}/n_{\rm H} = 0.085$ (number ratio). Identified photospheric lines are marked. Other labels are ``art.'' indicating instrument artefacts, ``i.s.'' interstellar lines, ``neb.'' nebular emission lines, ``sky'' night sky, and ``tell.'' telluric features. 
\label{fig:a57_publication}
}
\end{figure*}

For the analysis, we first normalized our LRS2-B spectrum of the central star to a flat continuum, shown as a blue line in Figure~\ref{fig:a57_publication}. We employed the T\"ubingen Model-Atmosphere Package \citep[TMAP;][]{Werner2003} to build a grid of non-LTE plane-parallel model atmospheres in radiative and hydrostatic equilibrium. 
Since no species heavier than helium are detected in the spectrum, we are unable to make any statement about the star's metal content. Consequently we considered model atmospheres composed only of hydrogen and helium.

We used the model spectra to determine the effective temperature, surface gravity, and He/H abundance ratio of the stellar atmosphere. The grid spacings for $\Teff$ and $\logg$ were 5000\,K and 0.2\,dex, respectively, while the He/H number ratio ranged from zero to 0.1 in initial steps of 0.01, with a finer spacing being used close to the finally determined value. For the best-fitting model, chosen by eye, we find $\Teff = 90,000\pm10,000$\,K, $\logg = 6.0\pm0.5$, and $n_{\rm He}/n_{\rm H} = 0.085\pm0.017$. The red line in Figure~\ref{fig:a57_publication} compares the model spectrum with the observed one, showing generally good agreement. The relatively large uncertainties in the stellar parameters are due to the fact that the photospheric Balmer line cores are filled in by nebular emissions from the CEK, so that the analysis is done using only the wings of the lines. Nebular emission also partially fills in \heii\  4686\,\AA\null. Thus, we determined the helium abundance only from the \heii\ lines at 4540 and 5410\,\AA\null. Within the uncertainties, the derived helium abundance is consistent with the solar value of ($n_{\rm He}/n_{\rm H })_\odot = 0.085$ \citep{Asplund+2009}. 


\subsection{Spectral-Energy Distribution and Reddening \label{sec:SED} }

As discussed in Section~\ref{sec:EGB6}, the nucleus of the prototypical EGB~6 emits an SED that combines the continuum of a hot central star with excesses at near-, mid-, and far-infrared wavelengths. We assembled the SED for the central star of Abell~57 in order to investigate whether it shares these phenomena.

To our knowledge, no previous ultraviolet (UV) observations of Abell~57 exist in the literature. Because such data are useful for constraining the extinction of the central star, we requested target-of-opportunity (TOO) imaging using the Ultraviolet Optical Telescope (UVOT; \citealt{Roming2005}) onboard the {\it Neil Gehrels Swift Observatory\/} \citep{Gehrels2004}. Conversion of the \Swift\/ count rates to monochromatic flux densities is a complicated process, which we describe in Appendix~\ref{appendix}, along with details of our TOO observations. 

Table~\ref{tab:a57_magnitudes} presents stellar magnitudes for Abell 57's central star in a range of bandpasses, and the corresponding absolute fluxes. These data were gathered from the following sources. (1)~UVOT magnitudes and fluxes from our new \Swift\/ observations, calibrated as described in Appendix~\ref{appendix}. (2)~Magnitudes (AB scale) in the $g$, $r$, $i$, $z$, and $y$ bands from the photometric catalog\footnote{\url{https://catalogs.mast.stsci.edu/panstarrs/}} of the Panoramic Survey Telescope and Rapid Response System (Pan-STARRS or PS1). (3)~The $u$-band magnitude from the Sloan Digital Sky Survey (SDSS)\null. The central star itself is not contained in the SDSS photometric catalog,\footnote{\url{https://skyserver.sdss.org/dr18/VisualTools/quickobj}} but photometry is available for a number of nearby stars of similar brightness. Accordingly, we downloaded an SDSS $u$-band image of the field\footnote{From the image server at \url{https://skyserver.sdss.org/dr18/VisualTools/}} and performed aperture photometry on the central star, along with several adjacent cataloged stars, using standard tasks in {\tt IRAF}.\footnote{{\tt IRAF} was distributed by the National Optical Astronomy Observatories, operated by AURA, Inc., under cooperative agreement with the National Science Foundation.} These measures were used to determine the central star's magnitude in the AB system. (4)~NIR magnitudes in the Vega-based $JHK_s$ system of the Two Micron All Sky Survey (2MASS; \citealt{Cohen-2MASS}). The faint nucleus itself is not contained in the 2MASS photometric catalog,\footnote{\url{https://irsa.ipac.caltech.edu/Missions/2mass.html}} but many brighter nearby stars are. We downloaded images of the field from the 2MASS Atlas Image Service,\footnote{\url{https://irsa.ipac.caltech.edu/applications/2MASS/IM/}} and carried out aperture photometry of the central star, and of several brighter neighbors for calibration. The nucleus was too faint to be detected in the $K_s$ band, but we successfully made measurements in $J$ and $H$\null.  (5)~MIR magnitudes in the Vega-based system of the {\it Wide-field Infrared Survey Explorer\/} \citep[\WISE;][]{Wright2010}, obtained from the NASA/IPAC Infrared Science Archive.\footnote{\url{http://irsa.ipac.caltech.edu/frontpage}} We also downloaded and inspected the \WISE\/ images from the science archive. The central star's cataloged measurement in the $W1$ band appears reliable, but in $W2$ there may be some contamination from a nearby field star, and the $W3$ and $W4$ images are clearly non-stellar, primarily due to dust in the surrounding PN\null. Accordingly we do not include these two magnitudes in our analysis.

\begin{deluxetable}{lcccc}
\tablecaption{Spectral-Energy Distribution of Abell 57 Central Star \label{tab:a57_magnitudes} } 
\tablewidth{0pt}
\tablehead{
\colhead{Bandpass}  &
\colhead{Magnitude}  &
\colhead{Source\tablenotemark{a}}  &
\colhead{$\lambda_{\rm eff}$\tablenotemark{b}}  &
\colhead{$F_\lambda$\tablenotemark{c}}  \\
\colhead{ }  &
\colhead{ }  &
\colhead{ }  &
\colhead{[$\mu$m]}  &
\colhead{[$\rm erg\,cm^{-2}\,s^{-1}\,$\AA$^{-1}$]}  
}  
\startdata 
$uvw2$ & 18.42\tablenotemark{d} & \Swift\/ & 0.203 & $6.81\times10^{-16}$ \\
$uvw1$ & 18.25\tablenotemark{d} & \Swift\/ & 0.260 & $9.05\times10^{-16}$ \\
\noalign{\vskip0.1in} 
$u$   & 17.848  & SDSS 	  & 0.356 & $6.24\times10^{-16}$  \\
$g$   & 17.686 & PS1 	  & 0.481 & $3.96\times10^{-16}$  \\
$r$   & 17.565 & PS1 	  & 0.617 & $2.69\times10^{-16}$  \\
$i$   & 17.630 & PS1 	  & 0.752 & $1.71\times10^{-16}$  \\
$z$   & 17.628 & PS1 	  & 0.866 & $1.29\times10^{-16}$  \\
$y$   & 17.604 & PS1 	  & 0.962 & $1.07\times10^{-16}$  \\
\noalign{\vskip0.1in} 
$J$   & 16.61 & 2MASS    & 1.241  & $7.66\times10^{-17}$  \\
$H$   & 16.17 & 2MASS    & 1.651  & $4.72\times10^{-17}$  \\
\noalign{\vskip0.1in} 
$W$1  & 14.970 & \WISE\/  & 3.37   & $1.13\times10^{-17}$  \\
$W$2  & 14.256 & \WISE\/  & 4.62   & $8.47\times10^{-18}$  \\
$W$3  & 9.612  & \WISE\/  & 12.08  & $1.01\times10^{-17}$  \\
$W$4  & 7.392  & \WISE\/  & 22.19  & $6.90\times10^{-18}$  \\
\enddata 
\tablenotetext{a}{Sources for magnitudes in column 2 are explained in the text. Uncertainties are about $\pm$0.03~mag (statistical) and $\pm$0.05~mag (systematic) for \Swift, $\pm$0.01-0.02~mag for the SDSS and PS1 magnitudes, $\sim$$\pm$0.1~mag for 2MASS, and approximately $\pm$0.05~mag for \WISE\/ $W1$ and $W2$. The $W3$ and $W4$ magnitudes are heavily contaminated by the surrounding PN. }
\tablenotetext{b}{Effective wavelengths for the \Swift\/ bandpasses are discussed in Appendix~\ref{appendix}. For SDSS $u$ the effective wavelength is from \citet{Fukugita1996}, and for PS1 they are from \citet{Tonry2012}. For 2MASS and \WISE\/ they are taken from the compilation at \url{http://coolwiki.ipac.caltech.edu/index.php/Central_wavelengths_and_zero_points}.}
\tablenotetext{c}{Absolute fluxes for \Swift\/ UVOT are discussed in Appendix~\ref{appendix}. For 2MASS and \WISE\/ the zero-points are from the compilation cited in Footnote~b. SDSS and PS1 magnitudes are on the AB scale.} 
\tablenotetext{d}{Nominal AB-scale magnitudes from \Swift\/ UVOT; conversion to absolute fluxes is described in Appendix~\ref{appendix}.}

\end{deluxetable}

\begin{figure}
\centering
\includegraphics[width=\linewidth]{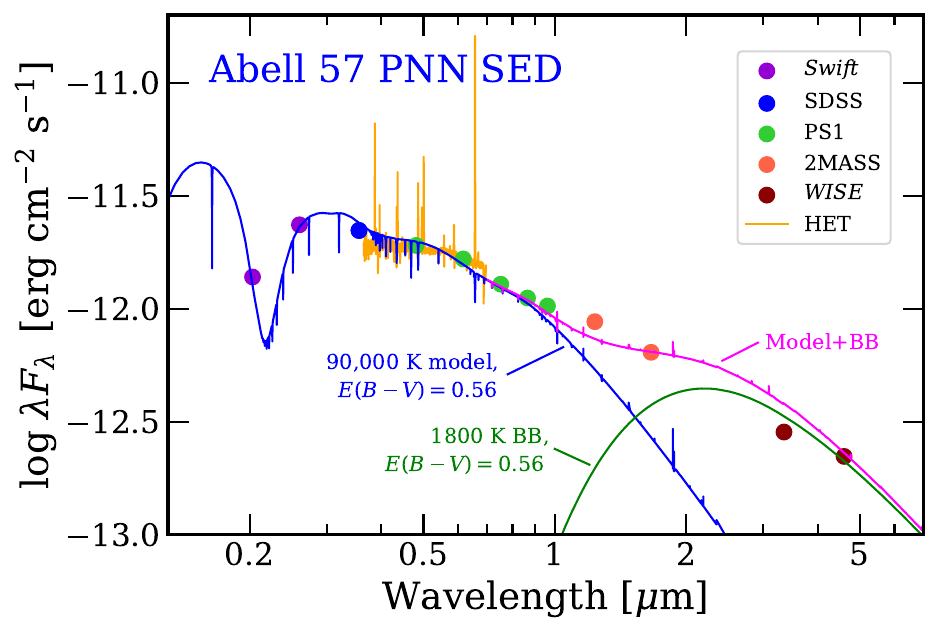}
\caption{
Spectral-energy distribution for the nucleus of Abell 57. Filled circles plot the photometric measurements from \Swift, SDSS, PS1, 2MASS, and \WISE\/ (see Table~\ref{tab:a57_magnitudes} for details; {  uncertainties are smaller than the plotting symbols, except for 2MASS, where they are approximately the size of the symbols}). The orange curve is the observed HET spectrum, and the blue curve shows the theoretical spectrum for the $(\Teff,\log g)=(90000,6.0)$ model described in Section~\ref{sec:atmosph_analysis}, with reddening of $E(B-V)=0.56$ applied. The green curve is an 1800~K blackbody, with the same reddening. The magenta curve plots the sum of the reddened model spectrum and the blackbody. A near-IR excess appears at wavelengths longward of $\sim$1~$\mu$m, similar to what is seen in EGB~6. The two reddest \WISE\/ W3 and W4 points from Table~\ref{tab:a57_magnitudes} are omitted, as they are likely due to the surrounding PN, rather than the central star. 
\label{fig:a57_sed}
}
\end{figure}

The final column in Table~\ref{tab:a57_magnitudes} gives the absolute fluxes for the central star, converted from the magnitudes in column~2 using the zero-points referenced in the table footnotes. These fluxes are plotted against wavelength in Figure~\ref{fig:a57_sed}. Also displayed as a blue curve is the synthetic spectrum from the TMAP model, normalized to match the observed fluxes at visual wavelengths. We reddened the model spectrum by various amounts, using the formulation of \citet{Cardelli1989}, in order to find the best fit to the observations. The UV fluxes in particular tightly constrain the reddening to a value of $E(B-V)=0.56$; {  changes of more than $\pm$0.02 produce a clear discrepancy with the model-atmosphere SED}\footnote{  As noted in Section~\ref{sec:atmosph_analysis}, the model atmosphere contains only H and He. Since the dominant opacity sources determining the SED are electron scattering and the bound-free and
free-free opacities of H and He, the lack of metals in the model does not affect
the reddening determination.} (assuming a fixed value of $R_V=3.1$). As the figure shows, the fluxes at wavelengths below about $1\,\mu$m are fit very well by the theoretical model reddened by this amount.\footnote{Our HET/LRS2 spectrum is plotted as an orange curve in Figure~\ref{fig:a57_sed}. It agrees very well with the PS1 fluxes (and with the reddened model SED), but it is slightly fainter than indicated by the SDSS $u$-band flux. Offsets of this size are consistent with the random errors found in our standard-star extractions, which are often largest in the LRS2's UV channel.}

The central star's reddening is inconsistent with predictions of the foreground extinction based on all-sky surveys. The Galactic dust reddening map of \cite{Schlafly2011}\footnote{As  implemented at \url{https://irsa.ipac.caltech.edu/applications/DUST/}} indicates a total reddening of $E(B-V)=0.37$ in the direction of Abell~57, and the online {\tt Stilism} tool\footnote{\url{https://stilism.obspm.fr/}} of \citet{Capitanio2017} gives $E(B-V)=0.33\pm0.04$ for a distance of 2100~pc. Both of these values are in fairly good agreement with, or slightly smaller than, the reddening inferred from the spectrum of the faint surrounding PN, as discussed below (Section~\ref{sec:PNspectrumanalysis}). Thus the observations imply that the central star suffers more extinction than expected from foreground dust.

At wavelengths longer than about $1\,\mu$m, the measured flux begins to exceed that expected from the 90,000~K central star.\footnote{\citet{DeMarco2013} reported that Abell~57 shows an excess at wavelengths as short as that of the $I$ band ($\lambda_{\rm eff}=7980$~\AA), but this is not confirmed by the PS1 data plotted in Figure~\ref{fig:a57_sed}.} This NIR excess can be represented reasonably well by a blackbody with a temperature of about 1800~K, plotted as a green curve in Figure~\ref{fig:a57_sed}. The sum of the fluxes of the O(H) nucleus and the cooler blackbody is shown as a magenta curve; it passes close to the four NIR points from 2MASS and \WISE\null. Thus Abell~57 is remarkably similar to the central star of EGB~6, which hosts a NIR source with an effective temperature of $\sim$1850~K (see Section~\ref{sec:EGB6} and B16). At a distance of $\sim$2100~pc, the 1800~K blackbody component of the Abell~57 nucleus has a luminosity of about $\log L/L_\odot = -1.1$. This is approximately the luminosity of an M0 dwarf.\footnote{Based on the compilation of stellar data assembled by E.~Mamajek, at \url{https://www.pas.rochester.edu/~emamajek/EEM_dwarf_UBVIJHK_colors_Teff.txt}} However, the temperature of the source is much cooler than that of an M0 dwarf, which is about 3850~K\null. Moreover, based on this component's luminosity and temperature, the implied radius of the NIR source is roughly $2.8\,R_\odot$, much larger than the radius of any low-mass main-sequence star.

{ 

In the case of EGB~6, \HST\/ imaging revealed the NIR source to be spatially resolved from the hot central star, and NIR spectra from the Gemini Observatory showed it to have a featureless continuum. The same may be true of the cool source in Abell~57, but high-resolution imaging and NIR spectroscopy are required for confirmation. 

}





\subsection{Evolutionary Status of the Central Star \label{sec:evolstatus} }

\begin{figure}
\centering
\includegraphics[width=\linewidth]{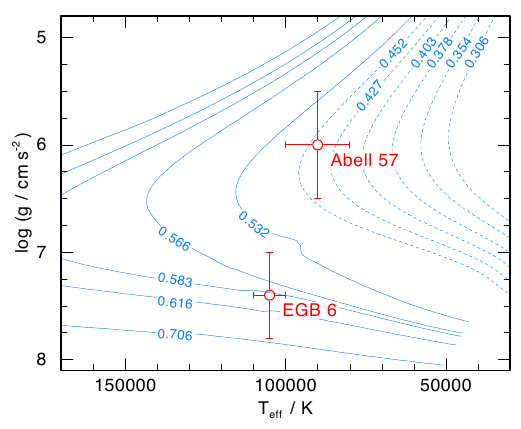}
\caption{
Position of the central stars of Abell~57 (this work) and EGB~6 \citep{WernerEGB62018} in the $\logg$--$\Teff$ diagram. Evolutionary tracks for post-AGB remnants (solid blue lines) by \citet{millerbertolami16} and for post-RGB remnants (dashed blue lines) by \citet{Hall2013} are labelled with the remnant stellar masses in solar units.
\label{fig:gteff}
}
\end{figure}

In Figure~\ref{fig:gteff} we show the position of the Abell~57 central star in the ``Kiel'' ($\logg$--$\Teff$) diagram. Also shown are theoretical post-asymptotic-giant-branch (post-AGB) evolutionary tracks (solar metallicity) from \cite{millerbertolami16}, plotted as solid blue lines, and post-red-giant-branch (post-RGB) tracks from \citet{Hall2013}, plotted as dashed blue lines. The star lies to the right of the lowest-mass post-AGB track, but within the uncertainties it could still be considered to be a post-AGB object. In this case, by linear extrapolation we determine the central star's mass to be $M = 0.50 \pm 0.06\,\msun$. The post-AGB age can also be determined from the evolutionary tracks; however, because of the strong dependence of the evolutionary speed on stellar mass and considering the errors of the spectroscopic analysis, the age constraint is very loose, $t_{\rm evol} = 52^{+4}_{-45}$\,kyr. In comparison, the evolutionary age determined for the central star of EGB~6 is $t_{\rm evol} = 4.0^{+68}_{-0.76}$\,kyr \citep{WernerEGB62018}. We conclude that we are unable to determine which of the two central stars has a greater post-AGB age. However, as noted in Section~\ref{sec:abell57}, the smaller physical size of the Abell~57 PN suggests that it is a younger object.

Support for a younger age of Abell~57 comes the kinematics of the nebula.  Assuming an expansion velocity of 24~km\,s$^{-1}$ \citep{Pereyra2013} the kinematic age of Abell~57 is about 8,000~yr.  For comparison, the expansion velocity of EGB~6 \citep[38~km\,s$^{-1}$;][]{Hippelein1990} yields an estimated age of 15,000~yr \citep{WernerEGB62018}. 

As noted, within the errors of the spectroscopic analysis, the stellar parameters of Abell~57 are consistent with a low-mass post-AGB star. However, an alternative possibility is that the nucleus actually is in a post-RGB stage (i.e., a star that underwent strong mass loss at the tip of the red-giant branch and failed to ignite helium-burning). If this were the case, the evolutionary tracks from \citet{Hall2013}, shown as dashed lines in Figure~\ref{fig:gteff}, would imply a mass of $M = 0.45^{+0.06}_{-0.03}\,\msun$ and an age of $t_{\rm evol} = 55^{+37}_{-8}$\,kyr. Since this greatly exceeds the kinematic age of the PN given above, we consider the post-RGB scenario unlikely.\footnote{Nevertheless, it is true, and puzzling, that a very few central stars of PNe almost certainly do lie on post-RGB tracks; see the discussions in \citet{Jones2023} and our Paper~II.}

As a test of our derived stellar parameters, we can calculate the spectroscopic distance of the Abell~57 central star from the relation
$$ d \, {\rm [pc]}= 7.11\times 10^{4} \sqrt{H_\nu\cdot M\cdot 10^{0.4V_0-\log g}}\ ,$$
which is based on the flux calibration by \citet{Heber1984}. {  The derived distance is especially sensitive to surface gravity, with a change of 1~dex in $\log g$ propagating into a factor of $\sim$3 in distance.} Here $H_\nu = 1.31\times10^{-3}\,\rm erg\,cm^{-2} \, s^{-1} \, Hz^{-1}$ is the Eddington flux of the 90,000~K model atmosphere at 5400\,\AA, $M$ is the stellar mass (in $M_\odot$) given above, $V_0=V-A_V$ is the dereddened visual magnitude, $V$ is the observed magnitude, and $A_V$ is the visual extinction. 
We converted the PS1 $g$ magnitude listed in Table~\ref{tab:a57_magnitudes} to Johnson $V$, using the relations given by \citet[][their Table~6]{Tonry2012},
giving $V=17.63$.\footnote{Photoelectric photometry by \citet{Abell1966} gave $V=17.66$, in very good agreement with the PS1 value. Almost 5 decades later, \citet{DeMarco2013} measured $V=17.73$ on two photometric nights, in poorer agreement. The latter authors also reported short-term photometric variability, but gave no details. \label{footnote:photometry} } Adopting $A_V=3.1 \, E(B-V)$, with $E(B-V)=0.56$, we derive a distance of $d=2739^{+2416}_{-1214}$~pc. This is in statistical agreement with the \Gaia\/ parallax distance ($2091^{+359}_{-271}$~pc; see Section~\ref{sec:abell57}). Also, using these values of $\Teff$, $H_\nu$, and the unreddened $V$ magnitude, with the \Gaia\/ distance, we find that the central star's luminosity is $\log L/L_\odot\simeq2.7$.







\section{Nebular Analysis \label{sec:nebular_analysis} }

In this section we investigate the properties of Abell 57's surrounding faint PN and of its CEK, based on analyses of their emission-line spectra.

\subsection{Abell 57 Planetary Nebula \label{sec:PNspectrumanalysis} }

As described in Section~\ref{sec:LRS2_spectroscopy}, we extracted a flux-calibrated spectrum of the faint PN within a $2''$-radius aperture placed near the location of highest surface brightness in the FOV of our LRS2-B observations, centered $4''$ away from the central star ($\sim$0.04~pc at the distance of the object). This spectrum is plotted in the bottom panel of Figure~\ref{fig:a57_spectra}, and the extraction aperture is shown as dashed black circles in Figure~\ref{fig:direct_images}. As discussed in Section~\ref{sec:LRS2_spectroscopy}, the PN spectrum is not sky-subtracted; this avoids any biasing effects of a spatially varying nebular spectrum, but it does suffer from the superposed night-sky spectrum. 

We measured absolute emission-line fluxes in the PN spectrum, using standard tasks in {\tt IRAF\null}. These fluxes are listed in column~3 of Table~\ref{tab:a57_pn_lines}, normalized to a value of ${\rm H}\beta=100$. 
Uncertainties in the line fluxes are generally a few percent, but are several times larger for blended and\slash or weak lines, as indicated by colons following the values. 




\begin{deluxetable}{lcDDD}
\tablecaption{Spectrum of Abell 57 Planetary Nebula\label{tab:a57_pn_lines} }
\tablewidth{0pt}
\tablehead{
\colhead{Species}  &
\colhead{$\lambda$}  &
\multicolumn2c{$I_{\rm rel}$\tablenotemark{a}} &
\multicolumn2c{$I_0$\tablenotemark{b}} &
\multicolumn2c{$I_{\rm model}$\tablenotemark{c}} \\
\colhead{}  &
\colhead{[\AA]}  &
\multicolumn2c{[H$\beta=100$]} &
\multicolumn2c{[H$\beta=100$]} &
\multicolumn2c{[H$\beta=100$]}
}
\decimals
\startdata
\null[\oii]\tablenotemark{d} & 3726-29&12.9:  &19.8:  & 16.6 \\
\null[\neiii]       &3869   &  59.2   &  87.2   & 91.2 \\
H$\zeta$+\hei\tablenotemark{e} &3889   & 14.3   &  18.0   & 19.8 \\
\null[\neiii]       &3967   &  10.4:  &  15.0:  & 27.7 \\
H$\epsilon$         &3970   &   9.1:  &  13.0:  & 16.8 \\
H$\delta$           &4101   &  17.7   &  24.0   & 26.3 \\
H$\gamma$           &4340   &  36.8   &  45.3   & 48.2 \\
\null[\oiii]        &4363   &   8.5   &  10.4   & 20.1 \\
\hei                &4471   &   2.9   &   3.4   &  0.4 \\
\heii               &4686   &  24.3   &  26.0   &  30.3 \\
\null[\ariv]        &4711   &   5.2   &   5.5   &  7.4 \\
\null[\ariv]        &4740   &   3.6   &   3.8   &  5.4 \\
H$\beta$\tablenotemark{f}             & 4861   & 100     & 100     & 100   \\
\null[\oiii]        &4959   & 335     & 324     & 340   \\
\null[\oiii]        &5007   &1035     & 984     & 1016	\\
\heii               &5411   &   2.6:  &   2.2:  &  2.7 \\
H$\alpha$           &6562   & 427     & 287     & 273   \\
\hei                &6678   &   4.4:  &   2.9:  &  2.3 \\
\null[\sii]         &6716   &   4.7:  &   3.0:   & 1.9 \\
\null[\sii]         &6731   &   3.4:  &   2.2:   & 1.5 \\
\enddata
\tablenotetext{a}{Measured emission-line intensity relative to H$\beta=100$.}
\tablenotetext{b}{Relative line intensity corrected for reddening of $E(B-V)=0.40$.}
\tablenotetext{c}{Predicted relative line intensity from model (see text).}
\tablenotetext{d}{Entries are for sum of [\oii] doublet.}
\tablenotetext{e}{Feature at $\lambda$3889 is a blend of H$\zeta$ and \hei\ $\lambda$3889. Entries are for their sum.}
\tablenotetext{f}{Measured absolute intensity of H$\beta$ is $5.34\times10^{-15}\rm erg\,cm^{-2}\,s^{-1}$.}
\end{deluxetable}

We can make approximate estimates of the nebula's physical parameters, including the  interstellar reddening, using line diagnostics described by \citet{Osterbrock+06}.
Given the age and low surface brightness of the Abell 57 nebula, we would expect its electron density, $n_e$, to be relatively low, and indeed, the [\sii] line diagnostic shows this to be the case. The ratio [\sii] $\lambda$6716/$\lambda$6731 is poorly measured, due to faintness and superposition on the night-sky spectrum. However, the ratio is near the low-density limit of the diagnostic, implying an electron density of about $n_e \simeq 100\, \rm cm^{-3}$.  At this density, {  and at a nominal electron temperature of 10,000~K,} the intrinsic Balmer-line ratios are expected to be approximately H$\alpha$/H$\beta = 2.86$, H$\gamma$/H$\beta = 0.466$, etc.\ \citep[see][]{Pengelly1964}.  
{  
Comparing these with the measured ratios yields a nebular reddening of $E(B-V) \simeq  0.40$ (assuming a \citealt{Cardelli1989} law with $R_V = 3.1$).
The uncertainty in this result, based on the scatter between the values obtained from the various Balmer ratios, and for a plausible range of temperatures, is about $\pm$0.03. The derived value
} 
agrees fairly well with the values of foreground interstellar reddening of $E(B-V)=0.37$ and $0.33\pm0.04$ from the statistical tools cited in Section~\ref{sec:SED}, but may suggest a small amount of internal reddening in the PN\null. The line fluxes corrected for this amount of reddening are given in the fourth column of Table~\ref{tab:a57_pn_lines}. 





To test this simple line-diagnostic analysis, we modeled Abell 57's nebula with the \cloudy\ photoionization code \citep{Ferland1998, Ferland2017}. We used version C23.00, described by \citet{Chatzikos2023}. 
We caution however that such an analysis is unlikely to be highly realistic, since
our spectral aperture extends over only a very small central part of the nebula, whereas \cloudy\ models the integrated spectrum of the entire nebula. 


We adopted several simplifying assumptions: spherical geometry with the ionizing star at the center, a uniform electron density of $n_e = 100$~cm$^{-3}$, and a ``PN'' abundance pattern of individual elements from \citet{aller1983chemical}. We set the inner radius of the model nebula to 0.04~pc. We adopted initial central-star parameters from Section~\ref{sec:atmosph_analysis} of $\Teff = 90,000$~K, and $L = 500\ L_{\odot}$. We then attempted to match the observed line strengths by varying the outer radius, the electron temperature as a function of radius, the metallicity, and allowing small changes in the stellar temperature.

\begin{figure}
    \centering
    \includegraphics[width=\linewidth]{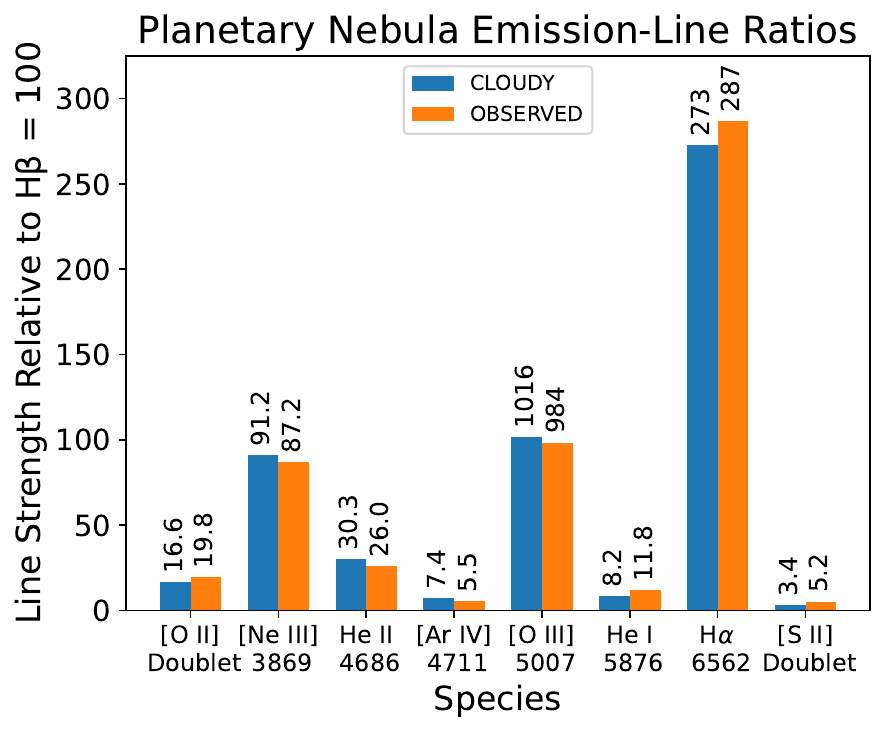}
    \caption{Emission-line fluxes in the Abell~57 planetary nebula, relative to H$\beta=100$, as observed corrected for extinction (orange bars) and obtained by our \cloudy\ model (blue bars). See text for details. The observed and modeled values for [\oiii] $\lambda$5007 have been scaled by a factor of 0.1 to improve visibility for the weaker lines.}
    \label{fig:pn_ratios}
\end{figure}

The best-fitting model was obtained with an electron temperature ranging from 17,030~K at the inner radius to 13,220~K at an outer radius of 0.66~pc. This large a radius arises from the need to reproduce the observed intensities of lower-ionization lines like [\oii] $\lambda$3726,3729. The adopted metallicity is 1/3 solar, and the effective temperature of the central star is adjusted slightly upward to 100,000~K\null. 
This model yields the emission-line fluxes given in the last column of Table~\ref{tab:a57_pn_lines}. In Figure \ref{fig:pn_ratios} we graphically compare the measured fluxes with the model predictions for several key emission lines, showing that the fluxes are predicted quite well---in spite of the \cloudy\ model being only a crude representation of the observation.

Our primary conclusion from this analysis is that the Abell~57 nebula is unexceptional, having a typical PN spectrum. It has an interstellar reddening of $E(B-V)\simeq0.40\pm0.03$, in reasonable agreement with the expected foreground amount. Of note, however, is the reddening of $E(B-V)=0.56\pm0.02$ that we found for the central star (Section~\ref{sec:SED}); {  it  is significantly} higher than it is in the surrounding faint PN.

\subsection{Compact Emission Knot}

As shown in the top panel of Figure~\ref{fig:a57_spectra}, the emission lines of Abell 57's CEK are superposed on the spectrum of the hot central star. We again measured the fluxes of the forbidden lines, using {\tt IRAF} tasks. However, the situation is complicated for the Balmer and helium lines, because they lie on top of underlying photospheric absorption lines instead of a featureless continuum. To measure the H and He line fluxes, we scaled the reddened theoretical spectrum from TMAP (Section~\ref{sec:atmosph_analysis}) to the level of the surrounding continuum, subtracted it from the observed spectrum, and then measured the line fluxes from that subtraction.  The resulting emission-line strengths, normalized to H$\beta=100$, are presented in column~3 of Table~\ref{tab:a57_knot_lines}. Weak and\slash or blended lines are again marked with colons. The lines of [\oii] $\lambda\lambda$3726,3729 are not detected, and instead we give an approximate upper limit on their combined flux.

\begin{deluxetable}{lcDDD}
\tablecaption{Spectrum of Abell 57 Compact Emission Knot\label{tab:a57_knot_lines} } 
\tablewidth{0pt}
\tablehead{
\colhead{Species}  &
\colhead{$\lambda$}  &
\multicolumn2c{$I_{\rm rel}$\tablenotemark{a}} &
\multicolumn2c{$I_0$\tablenotemark{b}} &
\multicolumn2c{$I_{\rm model}$\tablenotemark{c}} \\
\colhead{}  &
\colhead{[\AA]}  &
\multicolumn2c{[H$\beta=100$]} &
\multicolumn2c{[H$\beta=100$]} &
\multicolumn2c{[H$\beta=100$]}
} 
\decimals 
\startdata 
\null [\oii]\tablenotemark{d} &  3726-29 & <5.5 & <14.6 & 0.6 \\
\null{[\neiii]}   &  3869 &  113    & 273     & 208 \\
\null H$\zeta$+\hei\tablenotemark{e}     &  3889 &  11.7   & 27.8    & 29.8 \\
\null{[\neiii]}\tablenotemark{f}    &  3967 &  25.1:  & 56.3:   & 63.1 \\
\null H$\epsilon$\tablenotemark{f}   &  3970 &  6.6:  & 14.8:   & 17.8 \\
\null H$\delta$    &  4101 &  12.5   & 25.0    & 26.2 \\
\null H$\gamma$    &  4340 &  26.5   & 42.6    & 48.8 \\
\null{[\oiii]}    &  4363 &  49.5   & 77.8    & 93.8 \\
\null\heii\tablenotemark{g}        &  4686 &  15.7: & 15.9: & 4.9 \\
\null H$\beta$\tablenotemark{h}      &  4861 &  100    & 100 & 100 \\
\null{[\oiii]}    &  4959 &  39.5   & 36.6   & 57.8 \\
\null{[\oiii]}    &  5007 &  119    & 106   & 173 \\
\null\heii\tablenotemark{f}        &  5411 &  1.4: & 1.0: & 0.5 \\
\null\hei         &  5876 &  26.6   & 14.4   & 16.7 \\
\null{[\oi]}      &  6300 &   3.3:  & 1.5:   & 0.3 \\
\null H$\alpha$    &  6562 &  673    &  274  & 274 \\
\null{[\nii]}     &  6583  &  4.2:: & 1.7::   & 3.7 \\
\null\hei         &  6678 &  7.1::  &  2.8::  & 3.8 \\
\enddata 
\tablenotetext{a}{Measured emission-line intensity relative to H$\beta=100$.}
\tablenotetext{b}{Relative line intensity corrected for reddening of $E(B-V)=0.91$.}
\tablenotetext{c}{Predicted relative line intensity from model (see text).}
\tablenotetext{d}{Entries are for sum of [\oii] doublet.}
\tablenotetext{e}{Feature at $\lambda$3889 is a blend of H$\zeta$ and \hei\ $\lambda$3889. Entries are for their sum.}
\tablenotetext{f}{Fluxes uncertain due to blending and superposition on photospheric Balmer and interstellar \caii\ absorption.} 
\tablenotetext{g}{Fluxes uncertain due to superposition on photospheric \heii\ absorption.}
\tablenotetext{h}{Measured absolute intensity of H$\beta$ is $2.21\times10^{-15}\rm erg\,cm^{-2}\,s^{-1}$ (uncorrected for extinction).}
\end{deluxetable}

A remarkable feature of the CEK's spectrum is the strength of the auroral [\ion{O}{3}] line at 4363~\AA\null.  This line is generally considered to be a key diagnostic of nebular electron temperature \citep{Osterbrock+06}, and indeed measurements of it have been the goal of a number of \textsl{JWST\/} programs for direct metallicity determinations in the high-redshift universe \citep[e.g.,][]{Laseter2024}.  However, as illustrated in Figure~5.1 of \citet{Osterbrock+06}, [\ion{O}{3}] $\lambda$4363 is usually expected to be quite weak, with a strength ranging between $\sim$1/30 and $\sim$1/500 that of [\ion{O}{3}] $\lambda 5007$. In the spectrum of the Abell~57 PN, the $\lambda$4363 flux is consistent with this range, at about 1/100 that of $\lambda$5007. Such weakness is certainly not the case in Abell~57's unresolved knot: as the top panel of Figure~\ref{fig:a57_spectra} illustrates, [\ion{O}{3}] $\lambda 4363$ is actually stronger than [\ion{O}{3}] $\lambda$4959, and is even $\sim$1/2.5 the strength of $\lambda 5007$ before correction for reddening.  

The explanation for the strong [\ion{O}{3}] $\lambda 4363$ lies in the knot's electron density.  At densities above $\sim$$10^6\,\rm cm^{-3}$, the 
2p$^2$ $^1$D$_2$ level of O$^{++}$ becomes collisionally depopulated, weakening the nebular lines of [\ion{O}{3}] $\lambda\lambda 4959,5007$.  In contrast, the critical density for the parent level  of [\ion{O}{3}] $\lambda 4363$ ($^1$S$_0$) is not reached until $n_e \gtrsim 4 \times 10^7\,\rm cm^{-3}$.  Consequently, at high nebular densities, the ratio of the auroral to nebular lines of O$^{++}$ is boosted, and the line becomes a density diagnostic instead of a temperature indicator.  The CEK's high density also explains our non-detection of [\ion{O}{2}] $\lambda\lambda$3726,3729, whose critical density is only 1200~cm$^{-3}$. 

A second notable feature in the spectrum of the CEK is the large size of the Balmer decrement.  As described in the previous subsection, the observed H$\alpha$/H$\beta$ ratio for the Abell~57 PN is 4.27 (see~Table~\ref{tab:a57_pn_lines}), which we used to infer an interstellar reddening of $E(B-V)\simeq0.40$.  In the CEK the H$\alpha$/H$\beta$ ratio is much higher, at a value of 6.73 (Table~\ref{tab:a57_knot_lines}).  This implies dust extinction for the knot that is greater than for the surrounding PN (as well as being even higher than we found for the central star).

To fix the reddening of Abell 57's knot, we used the results of \citet{Storey-Hummer1995}, who tabulate the intrinsic recombination line strengths of hydrogenic ions in the range $5,000 \, {\rm K} \leq T_{\rm e} \leq 20,000\,\rm K$ and $0 \leq \log n_e \leq 8$ 
{  
(thus extending the density range covered by \citealt{Pengelly1964} and \citealt{Osterbrock+06} by 2~dex.)
}
The paucity of emission lines measured in the CEK prevents us from measuring the region's electron density and temperature simultaneously.  Because the efficiency of collisional cooling decreases with density, the electron temperature in the CEK is unlikely to be lower than 9,000~K\null.  Under these conditions, $n_e \approx 4 \times 10^7\,\rm cm^{-3}$; alternatively, if the knot's electron temperature is as high as $T_e \simeq 13,000$~K, then $n_e \approx 1.5\times10^7$~cm$^{-3}$.  In either case, the observed Balmer decrement and a \citet{Cardelli1989} reddening law with $R_V = 3.1$ give a reddening of $E(B-V) \simeq 0.91$. {  As in the case of the PN's spectrum, the uncertainty in the reddening is estimated to be about $\pm$0.03.} Note that the $V$-band extinction, $A_V=3.1\,E(B-V)$, in Abell 57's CEK is more than 1.5~mag greater than that of the surrounding nebula, and more than one magnitude higher than that found from the continuum of the central star.  The line fluxes corrected for this amount of extinction, again using the formulation of \citet{Cardelli1989}, are listed in column~4 of Table~\ref{tab:a57_knot_lines}.

\begin{figure}
    \centering
    \includegraphics[width=\linewidth]{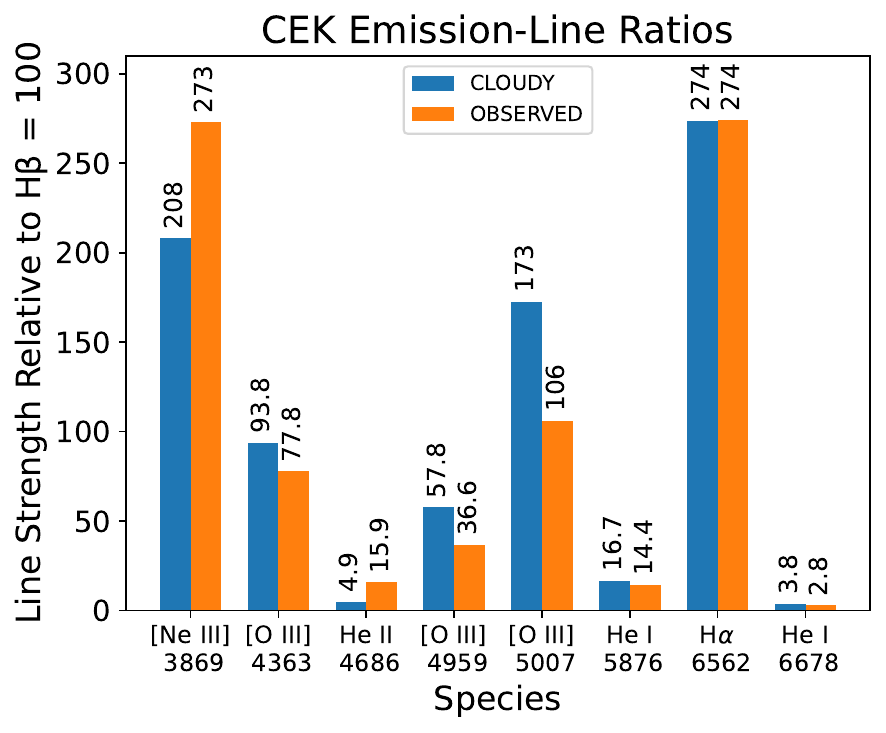} 
    \caption{Emission-line fluxes in the Abell~57 compact emission knot, relative to H$\beta=100$, as observed corrected for extinction (orange bars) and obtained by our \cloudy\ model (blue bars). See text for details.}
    \label{fig:cek_ratios}
\end{figure}

To test these conclusions about the CEK, we performed a \cloudy\ analysis similar to the one described in Section \ref{sec:PNspectrumanalysis}, with appropriate initial parameters. In this case, the analysis should be more realistic than it was for the PN, since we have an integrated spectrum for the entire CEK\null. Noting that in the EGB~6 system the CEK lies more than a projected 100~AU from the central star, we assumed that Abell~57 has a similar geometry. We adopted a CEK electron density fixed at $n_e = 10^{7.2} \rm\,cm^{-3}$ ($\sim$$1.6\times10^{7}$~cm$^{-3}$), a ``PN'' element abundance pattern (again from \citealt{aller1983chemical}), the SED and luminosity of the 90,000~K illuminating star described in Sections~\ref{sec:atmosph_analysis} and \ref{sec:evolstatus}, and a separation of 200~AU between the star and the inner radius of the nebula.

We then carried out multiple iterations of \cloudy, varying 
the thickness of the knot, and its electron density and metallicity, in order to find the best match of the modeled line ratios with the observed reddening-corrected values, as well as the absolute flux of H$\beta$ at a distance of 2100~pc. We found a consistent mismatch in the strengths of [\neiii] $\lambda$3869 and [\oiii] $\lambda$5007, with the former being consistently underpredicted and the latter overpredicted. As the relative strengths of [\neiii] and [\oiii] are insensitive to changes in the model parameters, we were unable to find a good fit for both species. We ended up splitting the difference. The line fluxes obtained from our final model are shown in column~5 of Table \ref{tab:a57_knot_lines}. A graphical comparison of the predicted and observed fluxes is presented in Figure \ref{fig:cek_ratios}. The agreements are reasonably good, except for [\neiii] and [\oiii], as just described. The agreement is also poor for \heii\ $\lambda$4686, but here the measurement is very uncertain, due to the line being superposed on a strong photospheric absorption line.


Our best-fit \cloudy\ model yields a nebular volume of $10^{42}\,\rm cm^3$, corresponding to a sphere with a radius of $10^{13.8}\rm\,cm$ or $\sim$4.5~AU\null.  The electron temperature is $T_e=12,500$~K, and the metallicity is 2 times solar. The high metallicity is required to fit the strengths of [\neiii] and [\oiii], {  but it should be regarded as uncertain because of the mismatch between the [\neiii] and [\oiii] lines described above.} Based on the adopted density and derived volume, the ionized mass of the knot is very small, about $1.7\times10^{-8}\,M_\odot$.




\section{Discussion and Future Work}

\subsection{Astrophysical Puzzles of Abell 57 and EGB 6 \label{sec:puzzles} }

The main conclusions of our investigation are as follows: (1)~Abell~57 is a low-surface-brightness, but otherwise unexceptional PN with a kinematic age of about 8,000~yr. (2)~The reddening of the faint PN is $E(B-V)\simeq0.40$, consistent with the expected foreground Galactic extinction and little to no internal extinction. (3)~The central star of Abell~57 has a spectral type of O(H), a solar He/H abundance, an effective temperature of 90,000~K, and a surface gravity of $\log g=6$. These parameters are consistent, within the uncertainties, with a star in a post-AGB evolutionary stage with a mass of about $0.50\,M_\odot$. (4)~The central star is reddened by $E(B-V)=0.56$, which is greater than the reddening of the surrounding PN\null. (5)~The nucleus exhibits a NIR excess, corresponding to an object with the luminosity of an M0 dwarf but with a blackbody temperature of $\sim$1800~K and a radius of $\sim$$2.8\,R_\odot$. (6)~The central star is also associated with a compact nebular knot, unresolved from it in our ground-based data. The knot has a high electron density of $n_e\simeq1.6\times10^7\rm\,cm^{-3}$, a physical radius of about 4.5~AU, and an ionized mass of $\sim$$1.7\times10^{-8}\,M_\odot$. (7)~The emission-line spectrum of the compact knot shows a reddening of $E(B-V)\simeq0.91$, which is considerably greater than the reddening of the central star---which itself is more reddened than the surrounding PN. 

Several of these features make Abell~57 strikingly similar to the prototypical EGB~6, described in Section~\ref{sec:EGB6} and in more detail by B16. In both objects there are CEKs associated with the hot central stars. The EGB~6 CEK, according to the analysis by \citet{Dopita1989}, has an electron density of $2.2\times10^6\rm\,cm^{-3}$, and an ionized mass of $7\times10^{-10}\,M_\odot$---which should be increased to $\sim$$2\times10^{-9}\,M_\odot$ because these authors assumed a distance of 460~pc in their analysis, whereas the \Gaia-based distance is 750~pc.
L13 derived a radius of 3.6~AU for the ionized knot in EGB~6, which should be increased to $\sim$4.7~AU, since they adopted a distance of 576~pc.
Compared to Abell~57, the EGB~6 CEK's density and mass are an order of magnitude lower, but the radii are remarkably similar. 


Both nuclei are also associated with NIR flux excesses, which appear to be due to dust-enshrouded low-mass companion stars. The effective temperatures of these companion dust envelopes are similar, both being about 1800--1850~K, and they have luminosities equivalent to those of main-sequence stars of spectral type M0 (Abell~57) or M3.5 (EGB~6).\footnote{As discussed by B16, the companion stars could in principle be WDs, but if so a fine-tuning of their luminosities is required.} The radius of the EGB~6 NIR companion is $\sim$$0.9\,R_\odot$, versus $\sim$$2.8\,R_\odot$ for the presumably younger Abell~57; this difference may be suggestive of slow evaporation and\slash or accretion of the dusty material onto the M~dwarfs, if the original conditions were similar for both objects. 


{  

As summarized in Section~\ref{sec:EGB6}, the B16 study of EGB~6 presented a scenario in which a low-mass companion to
a hot PNN captures an accretion disk during the primary's superwind stage, and that disk has survived
to the present day.
In EGB~6 the CEK coincides spatially with this companion, strongly suggesting that the emission knot consists of material being evaporated from the circumstellar material, and photoionized, by UV radiation from the nearby hot WD\null. In the absence of high-resolution imaging and NIR spectroscopy of Abell~57, we can only speculate that a similar scenario applies. 

}

However, there are also puzzles raised by significant differences between Abell~57 and EGB~6. The central star of the high-Galactic-latitude PN EGB~6 is nearly unreddened [B16 find $E(B-V)=0.02$], and its CEK also suffers essentially no extinction (see the discussion of its Balmer-line ratios in L13). However, as we showed in Section~\ref{sec:SED}, the central star of Abell~57 is {\it more\/} reddened than the surrounding faint PN [$E(B-V)=0.56$ for the central star, versus 0.40 for the PN]. Moreover, the CEK in Abell~57 has an {\it even higher\/} reddening of $E(B-V)=0.91$, determined using the same Balmer-line method as for EGB~6.

The high extinction of the emission-line spectrum of the Abell~57 CEK may be explained by
substantial dust being mixed up from the obscured NIR object into the extended emission-line region, thus increasing the amount of internal extinction in the surrounding ionized material. But this scenario does not account for the enhanced extinction of the central star itself. In the case of EGB~6, its NIR companion and associated CEK lie at least 125~AU away from the central star, and thus have little effect on it. A possible explanation for the reddening of the PNN in Abell~57 is that the CEK is so close to it that the star actually lies within the outer edge of the knot. Alternatively, if the orbital separation is larger, there conceivably could be a chance alignment such that the star shines through the outer layers of the knot as seen from our location. In either of these cases, we might expect orbital motion to cause the extinction of the star to vary, and thus its brightness. Unfortunately, because of the faintness of the central star, there is actually very little available information on its long-term photometric stability to test this hypothesis, apart from the scant data in Section~\ref{sec:evolstatus} and Footnote~\ref{footnote:photometry}. Yet another possibility is that the extinction of the WD in unrelated to that of the CEK, and is due instead to a separate cloud of cool dust around the central star.

\subsection{Why are EGB 6 Planetary Nuclei so Rare?}

As discussed in Section~\ref{sec:EGB6}, the ``EGB~6 phenomenon'' is extremely rare among PNNi. At first glance, this seems surprising. It is common for WDs to have companion stars at separations of a few to a few hundred AU \citep[see, e.g., the population-synthesis study by][]{Willems2004}, and these companions would have been embedded in a dense wind during PN ejection. There are known candidate progenitor objects, of which one of the best known is the Mira system, where accretion onto a companion at least 70~AU from the late-type primary's wind is directly observed (see L13, B16, and references therein). Likewise, there are binaries among PN central stars where accretion from an AGB wind onto the companion has clearly occurred in the recent past. These include barium stars like the nuclei of WeBo~1 \citep{Bond2003} and Abell~70 \citep{JonesA702022}, in which companions of the hot nuclei exhibit enhanced abundances of carbon and $s$-process elements accreted from an AGB outflow. There is also the class of ``Abell~35-type'' central stars (see Paper~IV and discussion therein), where late-type companions of the hot nuclei have been spun up to rapid rotation through accretion of matter and angular momentum. Yet we observe very few binary PNNi like EGB~6 and Abell~57 in which there are companions that are actually still embedded in conspicuous debris clouds with their outer layers photoionized by the nearby hot central stars. 

There is little known about the orbital periods of Abell~35-type binary nuclei, except that the period of LoTr~5 is about 7.4~yr, and for Abell~35 itself it is at least several decades (Paper~IV and references therein). However, for field barium stars (i.e., post-PN binary systems in which the nebula has dissipated long ago), there is substantial information about their orbital periods. A recent tabulation of the binary periods for 60 \baii\ stars by \citet{Escorza2023} shows that 90\% of them are less than 25~yr and 50\% are less than 5~yr, with a tail of longer periods extending up to roughly 80~yr. In contrast, the orbital period of the companion of EGB~6 is at least $\sim$1450~yr (B16). These considerations make it very unlikely that the very wide binary EGB~6 represents the progenitor population of the Abell~35 and barium stars. 

A possible explanation for the rarity of EGB~6-type nuclei could be that the lifetimes of the compact knots are very short. However, the kinematic ages of the Abell~57 and EGB~6 PNe (i.e., the times elapsed since their AGB winds were active) are $\sim$8,000~yr and $\sim$15,000~yr, respectively (see Section~\ref{sec:evolstatus})---yet both of them still have conspicuous CEKs. Thus, if the knots were indeed created during the AGB superwind phase, their lifetimes must be fairly long, comparable in fact to the lifetimes of the surrounding faint PNe. In that case, the EGB~6 phenomenon ought to be more common.


These considerations raise the possibility that our default scenario is incorrect, and that the CEKs in EGB~6-type PNNi are not debris clouds captured during PN formation. Instead they may have a separate origin requiring special circumstances, and likely having relatively short lifetimes. One possibility worth considering is that the EGB~6 phenomenon may be related to the IR excesses and debris disks that are detected around a significant fraction of single WDs. These objects are discussed in an extensive literature \citep[e.g.,][]{Chu2011, Farihi2016, Veras2021, Brouwers2022}. The disks are generally considered to result from planetary systems destabilized because of the sudden mass loss from the WD progenitor at the end of its AGB phase. This leads to collisions and tidal disruptions of planets, asteroids, and\slash or comets, creating dusty debris around the WDs. However, to our knowledge, forbidden-line emission has not been detected in these objects---in many cases the WDs are too cool to photoionize circumstellar gas---so the connection to EGB~6 PN nuclei remains unclear.



\subsection{Future Studies }

There are several avenues of investigation that might shed light on the puzzling nature of Abell~57 and the EGB~6-type PN central stars. Most important would be high-resolution direct imaging of Abell~57, both in emission lines like [\oiii] $\lambda$5007 and in the NIR, to determine whether there is a resolved companion as there is in EGB~6 itself. These observations would be challenging, but not impossible, with \HST\null. In the case of EGB~6, the NIR companion (and superposed CEK) lies only $0\farcs16$ from the WD central star---and Abell~57 is 2.8 times more distant from us. However, our discussion of the reddening of the nucleus of Abell~57 suggests that its CEK may be considerably closer to the central star than it is in EGB~6. Or, conceivably, Abell~57 could be different, with its CEK coinciding with the central star, rather than with the NIR companion. Precise astrometry with \HST\/ could make this test. Another important study would be long-term photometric monitoring of the central star, since, as noted in Section~\ref{sec:puzzles}, the extinction of the star may be varying with orbital motion.

NIR spectroscopy of Abell~57 would be extremely useful. In the case of EGB~6, the NIR shows a continuous spectrum (B16), consistent with the NIR flux being due to warm dust that obscures the companion star, presumably an M-type dwarf. We expect the same to be true of Abell~57, which would confirm its similarity to the prototype.

The central star of EGB~6 is a conspicuous source at 24~$\mu$m in the MIR \citep{Chu2011}. One of the puzzles of EGB~6 is that the precise spatial location of this MIR excess is unknown; it could be associated with the CEK, or it might coincide with the WD central star and be unrelated to the emission knot. See B16 for further discussion of this issue; as they point out, precision astrometry with \JWST\/ could distinguish between these two locations of the cool dust. In the case of Abell~57, the central star is embedded in a surrounding PN that is bright in the MIR (see Section~\ref{sec:SED} and the \WISE\/ images in the HASH database), so we do not as yet know whether the nucleus itself has a MIR excess. Here again high-resolution \JWST\/ MIR images would be decisive.



\acknowledgments

We thank the HET queue schedulers and nighttime observers at McDonald Observatory for obtaining the data discussed here.

A.S.C. and H.E.B. thank the Faculty Undergraduate Research Support program, sponsored by the Pennsylvania State University Eberly College of Science, Office of Science Engagement, for partial support of publication costs.

The Low-Resolution Spectrograph 2 (LRS2) was developed and funded by The University of Texas at Austin McDonald Observatory and Department of Astronomy, and by The Pennsylvania State University. We thank the Leibniz-Institut f\"ur Astrophysik Potsdam (AIP) and the Institut f\"ur Astrophysik G\"ottingen (IAG) for their contributions to the construction of the integral-field units.

We acknowledge the Texas Advanced Computing Center (TACC) at The University of Texas at Austin for providing high-performance computing, visualization, and storage resources that have contributed to the results reported within this paper.

This work has made use of data from the European Space Agency (ESA) mission
{\it Gaia\/} (\url{https://www.cosmos.esa.int/gaia}), processed by the {\it Gaia\/}
Data Processing and Analysis Consortium (DPAC,
\url{https://www.cosmos.esa.int/web/gaia/dpac/consortium}). Funding for the DPAC
has been provided by national institutions, in particular the institutions
participating in the {\it Gaia\/} Multilateral Agreement.

This research has made use of the SIMBAD database, operated at CDS, Strasbourg, France.

We acknowledge the use of public data from the \Swift\/ HEASARC data archive, and we thank the \Swift\/ team and PI for their quick acceptance of our Target of Opportunity request.

Funding for SDSS-III has been provided by the Alfred P. Sloan Foundation, the Participating Institutions, the National Science Foundation, and the U.S. Department of Energy Office of Science. The SDSS-III web site is \url{http://www.sdss3.org/}.
SDSS-III is managed by the Astrophysical Research Consortium for the Participating Institutions of the SDSS-III Collaboration including the University of Arizona, the Brazilian Participation Group, Brookhaven National Laboratory, Carnegie Mellon University, University of Florida, the French Participation Group, the German Participation Group, Harvard University, the Instituto de Astrofisica de Canarias, the Michigan State/Notre Dame/JINA Participation Group, Johns Hopkins University, Lawrence Berkeley National Laboratory, Max Planck Institute for Astrophysics, Max Planck Institute for Extraterrestrial Physics, New Mexico State University, New York University, Ohio State University, Pennsylvania State University, University of Portsmouth, Princeton University, the Spanish Participation Group, University of Tokyo, University of Utah, Vanderbilt University, University of Virginia, University of Washington, and Yale University. 

The Pan-STARRS1 Surveys (PS1) and the PS1 public science archive have been made possible through contributions by the Institute for Astronomy, the University of Hawaii, the Pan-STARRS Project Office, the Max-Planck Society and its participating institutes, the Max Planck Institute for Astronomy, Heidelberg and the Max Planck Institute for Extraterrestrial Physics, Garching, The Johns Hopkins University, Durham University, the University of Edinburgh, the Queen's University Belfast, the Harvard-Smithsonian Center for Astrophysics, the Las Cumbres Observatory Global Telescope Network Incorporated, the National Central University of Taiwan, the Space Telescope Science Institute, the National Aeronautics and Space Administration under Grant No.\ NNX08AR22G issued through the Planetary Science Division of the NASA Science Mission Directorate, the National Science Foundation Grant No. AST-1238877, the University of Maryland, Eotvos Lorand University (ELTE), the Los Alamos National Laboratory, and the Gordon and Betty Moore Foundation.

This research has made use of the NASA/IPAC Infrared Science Archive, which is operated by the Jet Propulsion Laboratory, California Institute of Technology, under contract with NASA.

This publication makes use of data products from the Two Micron All Sky Survey, which is a joint project of the University of Massachusetts and the Infrared Processing and Analysis Center/California Institute of Technology, funded by NASA and the NSF.

It also makes use of data products from the {\it Wide-field Infrared Survey Explorer}, which is a joint project of the University of California, Los Angeles, and the Jet Propulsion Laboratory/California Institute of Technology, and NEOWISE, which is a project of the Jet Propulsion Laboratory/ California Institute of Technology. \WISE\/ and NEOWISE are funded by NASA.

The Institute for Gravitation and the Cosmos is supported by the Eberly College of Science and the Office of the Senior Vice President for Research at The Pennsylvania State University.

\appendix

\section{\emph{Swift} Observations of Abell 57 and Data Analysis \label{appendix} }

As described in Section~\ref{sec:SED}, near-UV images of Abell~57 were obtained with the UVOT instrument onboard the {\it Neil Gehrels Swift Observatory}. In this Appendix, we describe the conversion of source count rates from the UVOT frames into monochromatic flux densities. 
Our frames were obtained in the broadband $uvw1$ and $uvw2$ filters, which bracket the interstellar absorption feature at 2175~\AA\null. Details of our \Swift\/ TOO observations are given in Table~\ref{tab:swift}.

\begin{deluxetable}{lcc}[h]
\tablecaption{Log of \Swift\/ UVOT Observations of Abell 57 \label{tab:swift} }
\tablehead{
\colhead{Date} &
\colhead{Filter} &
\colhead{Exposure}\\
\colhead{[YYYY-MM-DD]} &
\colhead{} &
\colhead{[s]}}
\decimals
\startdata
2023-10-31 & {\it uvw1\/} & 1535.5 \\
2023-10-31 & {\it uvw2\/} & 1568.0 \\
2023-11-03 & {\it uvw1\/} & 512.9 \\
2023-11-03 & {\it uvw2\/} & 513.6\\
\enddata
\end{deluxetable}

We generated source count rates for the Abell 57 central star from the images, using the standard {\tt HEASoft} tool {\tt uvotsource},\footnote{\url{https://heasarc.gsfc.nasa.gov/lheasoft/ftools/}} which accounts for zero-point, coincidence loss, sensitivity decline, and flat-field variations \citep[see][]{Poole2008, Breeveld2011}. These rates have a statistical uncertainty of about $\pm$3\%. However, our data were taken while the \Swift\/ spacecraft was experiencing a guidance anomaly that caused UVOT images to be elongated beyond the nominal 2\farcs5 FWHM PSF\null. To account for this, we applied a correction factor of 1.1 to the count rates from {\tt uvotsource}, a preliminary value based on flux ratios between 7\farcs0 and 10\farcs0 apertures. This issue is still being investigated by the UVOT team, but we estimate that our photometry has an additional systematic uncertainty of $\pm$5\% due to this issue. The measured count rates for Abell 57, with this correction applied, are listed in column~3 of Table~\ref{tab:uvot}.

\begin{deluxetable}{lcccc}[h]
\tablecaption{UVOT Monochromatic Fluxes for Abell 57 \label{tab:uvot} }
\tablehead{
\colhead{Filter} &
\colhead{Effective} &
\colhead{Corrected Count} &
\colhead{Conversion} &
\colhead{$F_\lambda$}\\
\colhead{} &
\colhead{Wavelength [\AA]} &
\colhead{Rate [cts s$^{-1}$]} &
\colhead{Factor} &
\colhead{[$\rm erg\,cm^{-2}\,s^{-1}\,$\AA$^{-1}$]}  
}
\decimals
\startdata
\uvwtwo\/ & 2030 & 1.894 & $3.594\times10^{-16}$ & $6.81\times10^{-16}$ \\
\uvwone\/ & 2600 & 1.904 & $4.753\times10^{-16}$ & $9.05\times10^{-16}$ \\
\enddata
\end{deluxetable}

Conversion of count rates to a monochromatic flux density is not straightforward with the UVOT {\it uvw1\/} and {\it uvw2\/} filters, because both of them have a significant amount of transmission at wavelengths longward of their main near-UV bandpasses (``red tails'' or ``red leaks''). This issue has been addressed by \citet{Brown2010, Brown2016}. They adopt the photon-weighted effective wavelengths of these filters for an object with a Vega-like spectrum from \citet{Poole2008} as follows: 2030~\AA\ for \uvwtwo\/ and 2600~\AA\ for \uvwone. They then determined the ``conversion factors'' by which to multiply the count rates for a Vega-like star to yield the monochromatic fluxes at these effective wavelengths. For Vega, \citet{Brown2016} derived conversion factors in \uvwtwo\ and \uvwone\ of $6.028\times10^{-16}$ and $4.024\times10^{-16}\,\rm erg\,cm^{-2}\,s^{-1}\,$\AA$^{-1}/\rm(cts\,s^{-1})$, respectively. 

The conversion factors will of course be different for objects whose SEDs differ from that of Vega. \citet{Brown2010, Brown2016} calculated the conversion factors for a variety of unreddened sources with known SEDs, such as blackbodies and stars of known spectral types. The factors were determined by dividing the average flux density in a 50~\AA\ window centered on the effective wavelength by the count rate of the spectrum through that filter, and normalizing the result to that for Vega. 

Since Abell 57 is significantly reddened, we could not use the tables of conversion factors in the above references. Fortunately, we know the intrinsic SED of the star from our spectroscopic analysis (Section~\ref{sec:atmosph_analysis}). We began by applying a nominal amount of reddening to the model SED, using the \citet{Cardelli1989} formulation. We then calculated the fractions of the total photon counts in each UVOT filter that are due to 50~\AA\ windows at the effective wavelengths, by convolving the reddened SED with the UVOT system throughput tables\footnote{\url{https://heasarc.gsfc.nasa.gov/docs/heasarc/caldb/swift/docs/uvot/index.html}} of \citet{Breeveld2011}. For normalization, we compared these fractions with those for a similar convolution of the UVOT throughputs with the SED of Vega\footnote{File alpha\_lyr\_stis\_011.fits, available at \url{https://archive.stsci.edu/hlsps/reference-atlases/cdbs/current_calspec}} from \citet{Bohlin2020}.

We compared the resulting absolute near-UV fluxes with the reddened model SED for the central star. On this basis, we adjusted the assumed reddening, and repeated the calculation. After a few iterations, the solution converged upon a reddening of $E(B-V)=0.56$. The final conversion factors for Abell~57 with this amount of reddening are given in column~4 of Table~\ref{tab:uvot}, and the resulting absolute monochromatic fluxes in column~5. These values are entered in Table~\ref{tab:a57_magnitudes} in the main text.




\clearpage

\bibliography{PNNisurvey_refs}

\begin{thebibliography}{}
\expandafter\ifx\csname natexlab\endcsname\relax\def\natexlab#1{#1}\fi
\providecommand{\url}[1]{\href{#1}{#1}}

\bibitem[{{Abell}(1966)}]{Abell1966}
{Abell}, G.~O. 1966, \apj, 144, 259

\bibitem[{Aller \& Czyzak(1983)}]{aller1983chemical}
Aller, L.~H., \& Czyzak, S.~J. 1983, Astrophysical Journal Supplement Series, vol. 51, Feb. 1983, p. 211-247., 51, 211

\bibitem[{{Asplund} {et~al.}(2009){Asplund}, {Grevesse}, {Sauval}, \& {Scott}}]{Asplund+2009}
{Asplund}, M., {Grevesse}, N., {Sauval}, A.~J., \& {Scott}, P. 2009, \araa, 47, 481

\bibitem[{{Bailer-Jones} {et~al.}(2021){Bailer-Jones}, {Rybizki}, {Fouesneau}, {Demleitner}, \& {Andrae}}]{BailerJones2021}
{Bailer-Jones}, C.~A.~L., {Rybizki}, J., {Fouesneau}, M., {Demleitner}, M., \& {Andrae}, R. 2021, \aj, 161, 147

\bibitem[{{Bohlin} {et~al.}(2020){Bohlin}, {Hubeny}, \& {Rauch}}]{Bohlin2020}
{Bohlin}, R.~C., {Hubeny}, I., \& {Rauch}, T. 2020, \aj, 160, 21

\bibitem[{{Bond} {et~al.}(2016){Bond}, {Ciardullo}, {Esplin}, {Hawley}, {Liebert}, \& {Munari}}]{BondEGB62016}
{Bond}, H.~E., {Ciardullo}, R., {Esplin}, T.~L., {et~al.} 2016, \apj, 826, 139, (B16)

\bibitem[{{Bond} {et~al.}(2003){Bond}, {Pollacco}, \& {Webbink}}]{Bond2003}
{Bond}, H.~E., {Pollacco}, D.~L., \& {Webbink}, R.~F. 2003, \aj, 125, 260

\bibitem[{{Bond} {et~al.}(2023{\natexlab{a}}){Bond}, {Werner}, {Jacoby}, \& {Zeimann}}]{Bond2023a}
{Bond}, H.~E., {Werner}, K., {Jacoby}, G.~H., \& {Zeimann}, G.~R. 2023{\natexlab{a}}, \mnras, 521, 668, (Paper I)

\bibitem[{{Bond} {et~al.}(2023{\natexlab{b}}){Bond}, {Werner}, {Zeimann}, \& {Talbot}}]{Bond2023b}
{Bond}, H.~E., {Werner}, K., {Zeimann}, G.~R., \& {Talbot}, J. 2023{\natexlab{b}}, \mnras, 523, 3699, (Paper II)

\bibitem[{{Bond} \& {Zeimann}(2024)}]{Bond_Pa27_2024}
{Bond}, H.~E., \& {Zeimann}, G.~R. 2024, arXiv e-prints, arXiv:2404.07152, (Paper IV)

\bibitem[{{Breeveld} {et~al.}(2011){Breeveld}, {Landsman}, {Holland}, {Roming}, {Kuin}, \& {Page}}]{Breeveld2011}
{Breeveld}, A.~A., {Landsman}, W., {Holland}, S.~T., {et~al.} 2011, in American Institute of Physics Conference Series, Vol. 1358, Gamma Ray Bursts 2010, ed. J.~E. {McEnery}, J.~L. {Racusin}, \& N.~{Gehrels}, 373--376

\bibitem[{{Brouwers} {et~al.}(2022){Brouwers}, {Bonsor}, \& {Malamud}}]{Brouwers2022}
{Brouwers}, M.~G., {Bonsor}, A., \& {Malamud}, U. 2022, \mnras, 509, 2404

\bibitem[{{Brown} {et~al.}(2016){Brown}, {Breeveld}, {Roming}, \& {Siegel}}]{Brown2016}
{Brown}, P.~J., {Breeveld}, A., {Roming}, P. W.~A., \& {Siegel}, M. 2016, \aj, 152, 102

\bibitem[{{Brown} {et~al.}(2010){Brown}, {Roming}, {Milne}, {Bufano}, {Ciardullo}, {Elias-Rosa}, {Filippenko}, {Foley}, {Gehrels}, {Gronwall}, {Hicken}, {Holland}, {Hoversten}, {Immler}, {Kirshner}, {Li}, {Mazzali}, {Phillips}, {Pritchard}, {Still}, {Turatto}, \& {Vanden Berk}}]{Brown2010}
{Brown}, P.~J., {Roming}, P. W.~A., {Milne}, P., {et~al.} 2010, \apj, 721, 1608

\bibitem[{{Capitanio} {et~al.}(2017){Capitanio}, {Lallement}, {Vergely}, {Elyajouri}, \& {Monreal-Ibero}}]{Capitanio2017}
{Capitanio}, L., {Lallement}, R., {Vergely}, J.~L., {Elyajouri}, M., \& {Monreal-Ibero}, A. 2017, \aap, 606, A65

\bibitem[{{Cardelli} {et~al.}(1989){Cardelli}, {Clayton}, \& {Mathis}}]{Cardelli1989}
{Cardelli}, J.~A., {Clayton}, G.~C., \& {Mathis}, J.~S. 1989, \apj, 345, 245

\bibitem[{{Chatzikos} {et~al.}(2023){Chatzikos}, {Bianchi}, {Camilloni}, {Chakraborty}, {Gunasekera}, {Guzm{\'a}n}, {Milby}, {Sarkar}, {Shaw}, {van Hoof}, \& {Ferland}}]{Chatzikos2023}
{Chatzikos}, M., {Bianchi}, S., {Camilloni}, F., {et~al.} 2023, \rmxaa, 59, 327

\bibitem[{{Chonis} {et~al.}(2016){Chonis}, {Hill}, {Lee}, {Tuttle}, {Vattiat}, {Drory}, {Indahl}, {Peterson}, \& {Ramsey}}]{Chonis2016}
{Chonis}, T.~S., {Hill}, G.~J., {Lee}, H., {et~al.} 2016, in Society of Photo-Optical Instrumentation Engineers (SPIE) Conference Series, Vol. 9908, Ground-based and Airborne Instrumentation for Astronomy VI, ed. C.~J. {Evans}, L.~{Simard}, \& H.~{Takami}, 99084C

\bibitem[{{Chu} {et~al.}(2011){Chu}, {Su}, {Bilikova}, {Gruendl}, {De Marco}, {Guerrero}, {Updike}, {Volk}, \& {Rauch}}]{Chu2011}
{Chu}, Y.-H., {Su}, K. Y.~L., {Bilikova}, J., {et~al.} 2011, \aj, 142, 75

\bibitem[{{Cohen} {et~al.}(2003){Cohen}, {Wheaton}, \& {Megeath}}]{Cohen-2MASS}
{Cohen}, M., {Wheaton}, W.~A., \& {Megeath}, S.~T. 2003, \aj, 126, 1090

\bibitem[{{De Marco} {et~al.}(2013){De Marco}, {Passy}, {Frew}, {Moe}, \& {Jacoby}}]{DeMarco2013}
{De Marco}, O., {Passy}, J.-C., {Frew}, D.~J., {Moe}, M., \& {Jacoby}, G.~H. 2013, \mnras, 428, 2118

\bibitem[{{Dopita} \& {Liebert}(1989)}]{Dopita1989}
{Dopita}, M.~A., \& {Liebert}, J. 1989, \apj, 347, 910

\bibitem[{{Ellis} {et~al.}(1984){Ellis}, {Grayson}, \& {Bond}}]{Ellis1984}
{Ellis}, G.~L., {Grayson}, E.~T., \& {Bond}, H.~E. 1984, \pasp, 96, 283

\bibitem[{{Escorza} \& {De Rosa}(2023)}]{Escorza2023}
{Escorza}, A., \& {De Rosa}, R.~J. 2023, \aap, 671, A97

\bibitem[{{Farihi}(2016)}]{Farihi2016}
{Farihi}, J. 2016, \nar, 71, 9

\bibitem[{{Ferland} {et~al.}(1998){Ferland}, {Korista}, {Verner}, {Ferguson}, {Kingdon}, \& {Verner}}]{Ferland1998}
{Ferland}, G.~J., {Korista}, K.~T., {Verner}, D.~A., {et~al.} 1998, \pasp, 110, 761

\bibitem[{{Ferland} {et~al.}(2017){Ferland}, {Chatzikos}, {Guzm{\'a}n}, {Lykins}, {van Hoof}, {Williams}, {Abel}, {Badnell}, {Keenan}, {Porter}, \& {Stancil}}]{Ferland2017}
{Ferland}, G.~J., {Chatzikos}, M., {Guzm{\'a}n}, F., {et~al.} 2017, \rmxaa, 53, 385

\bibitem[{{Frew} \& {Parker}(2010)}]{FrewParker2010}
{Frew}, D.~J., \& {Parker}, Q.~A. 2010, \pasa, 27, 129

\bibitem[{{Fukugita} {et~al.}(1996){Fukugita}, {Ichikawa}, {Gunn}, {Doi}, {Shimasaku}, \& {Schneider}}]{Fukugita1996}
{Fukugita}, M., {Ichikawa}, T., {Gunn}, J.~E., {et~al.} 1996, \aj, 111, 1748

\bibitem[{{Fulbright} \& {Liebert}(1993)}]{Fulbright1993}
{Fulbright}, M.~S., \& {Liebert}, J. 1993, \apj, 410, 275

\bibitem[{{Gaia Collaboration} {et~al.}(2016){Gaia Collaboration}, {Prusti}, {de Bruijne}, {Brown}, {Vallenari}, {Babusiaux}, {Bailer-Jones}, {Bastian}, {Biermann}, {Evans}, {Eyer}, {Jansen}, {Jordi}, {Klioner}, {Lammers}, {Lindegren}, {Luri}, {Mignard}, {Milligan}, {Panem}, {Poinsignon}, {Pourbaix}, {Randich}, {Sarri}, {Sartoretti}, {Siddiqui}, {Soubiran}, {Valette}, {van Leeuwen}, {Walton}, {Aerts}, {Arenou}, {Cropper}, {Drimmel}, {H{\o}g}, {Katz}, {Lattanzi}, {O'Mullane}, {Grebel}, {Holland}, {Huc}, {Passot}, {Bramante}, {Cacciari}, {Casta{\~n}eda}, {Chaoul}, {Cheek}, {De Angeli}, {Fabricius}, {Guerra}, {Hern{\'a}ndez}, {Jean-Antoine-Piccolo}, {Masana}, {Messineo}, {Mowlavi}, {Nienartowicz}, {Ord{\'o}{\~n}ez-Blanco}, {Panuzzo}, {Portell}, {Richards}, {Riello}, {Seabroke}, {Tanga}, {Th{\'e}venin}, {Torra}, {Els}, {Gracia-Abril}, {Comoretto}, {Garcia-Reinaldos}, {Lock}, {Mercier}, {Altmann}, {Andrae}, {Astraatmadja}, {Bellas-Velidis}, {Benson}, {Berthier}, {Blomme}, {Busso}, {Carry}, {Cellino}, {Clementini},
  {Cowell}, {Creevey}, {Cuypers}, {Davidson}, {De Ridder}, {de Torres}, {Delchambre}, {Dell'Oro}, {Ducourant}, {Fr{\'e}mat}, {Garc{\'\i}a-Torres}, {Gosset}, {Halbwachs}, {Hambly}, {Harrison}, {Hauser}, {Hestroffer}, {Hodgkin}, {Huckle}, {Hutton}, {Jasniewicz}, {Jordan}, {Kontizas}, {Korn}, {Lanzafame}, {Manteiga}, {Moitinho}, {Muinonen}, {Osinde}, {Pancino}, {Pauwels}, {Petit}, {Recio-Blanco}, {Robin}, {Sarro}, {Siopis}, {Smith}, {Smith}, {Sozzetti}, {Thuillot}, {van Reeven}, {Viala}, {Abbas}, {Abreu Aramburu}, {Accart}, {Aguado}, {Allan}, {Allasia}, {Altavilla}, {{\'A}lvarez}, {Alves}, {Anderson}, {Andrei}, {Anglada Varela}, {Antiche}, {Antoja}, {Ant{\'o}n}, {Arcay}, {Atzei}, {Ayache}, {Bach}, {Baker}, {Balaguer-N{\'u}{\~n}ez}, {Barache}, {Barata}, {Barbier}, {Barblan}, {Baroni}, {Barrado y Navascu{\'e}s}, {Barros}, {Barstow}, {Becciani}, {Bellazzini}, {Bellei}, {Bello Garc{\'\i}a}, {Belokurov}, {Bendjoya}, {Berihuete}, {Bianchi}, {Bienaym{\'e}}, {Billebaud}, {Blagorodnova}, {Blanco-Cuaresma}, {Boch},
  {Bombrun}, {Borrachero}, {Bouquillon}, {Bourda}, {Bouy}, {Bragaglia}, {Breddels}, {Brouillet}, {Br{\"u}semeister}, {Bucciarelli}, {Budnik}, {Burgess}, {Burgon}, {Burlacu}, {Busonero}, {Buzzi}, {Caffau}, {Cambras}, {Campbell}, {Cancelliere}, {Cantat-Gaudin}, {Carlucci}, {Carrasco}, {Castellani}, {Charlot}, {Charnas}, {Charvet}, {Chassat}, {Chiavassa}, {Clotet}, {Cocozza}, {Collins}, {Collins}, {Costigan}, {Crifo}, {Cross}, {Crosta}, {Crowley}, {Dafonte}, {Damerdji}, {Dapergolas}, {David}, {David}, {De Cat}, {de Felice}, {de Laverny}, {De Luise}, {De March}, {de Martino}, {de Souza}, {Debosscher}, {del Pozo}, {Delbo}, {Delgado}, {Delgado}, {di Marco}, {Di Matteo}, {Diakite}, {Distefano}, {Dolding}, {Dos Anjos}, {Drazinos}, {Dur{\'a}n}, {Dzigan}, {Ecale}, {Edvardsson}, {Enke}, {Erdmann}, {Escolar}, {Espina}, {Evans}, {Eynard Bontemps}, {Fabre}, {Fabrizio}, {Faigler}, {Falc{\~a}o}, {Farr{\`a}s Casas}, {Faye}, {Federici}, {Fedorets}, {Fern{\'a}ndez-Hern{\'a}ndez}, {Fernique}, {Fienga}, {Figueras}, {Filippi},
  {Findeisen}, {Fonti}, {Fouesneau}, {Fraile}, {Fraser}, {Fuchs}, {Furnell}, {Gai}, {Galleti}, {Galluccio}, {Garabato}, {Garc{\'\i}a-Sedano}, {Gar{\'e}}, {Garofalo}, {Garralda}, {Gavras}, {Gerssen}, {Geyer}, {Gilmore}, {Girona}, {Giuffrida}, {Gomes}, {Gonz{\'a}lez-Marcos}, {Gonz{\'a}lez-N{\'u}{\~n}ez}, {Gonz{\'a}lez-Vidal}, {Granvik}, {Guerrier}, {Guillout}, {Guiraud}, {G{\'u}rpide}, {Guti{\'e}rrez-S{\'a}nchez}, {Guy}, {Haigron}, {Hatzidimitriou}, {Haywood}, {Heiter}, {Helmi}, {Hobbs}, {Hofmann}, {Holl}, {Holland}, {Hunt}, {Hypki}, {Icardi}, {Irwin}, {Jevardat de Fombelle}, {Jofr{\'e}}, {Jonker}, {Jorissen}, {Julbe}, {Karampelas}, {Kochoska}, {Kohley}, {Kolenberg}, {Kontizas}, {Koposov}, {Kordopatis}, {Koubsky}, {Kowalczyk}, {Krone-Martins}, {Kudryashova}, {Kull}, {Bachchan}, {Lacoste-Seris}, {Lanza}, {Lavigne}, {Le Poncin-Lafitte}, {Lebreton}, {Lebzelter}, {Leccia}, {Leclerc}, {Lecoeur-Taibi}, {Lemaitre}, {Lenhardt}, {Leroux}, {Liao}, {Licata}, {Lindstr{\o}m}, {Lister}, {Livanou}, {Lobel}, {L{\"o}ffler},
  {L{\'o}pez}, {Lopez-Lozano}, {Lorenz}, {Loureiro}, {MacDonald}, {Magalh{\~a}es Fernandes}, {Managau}, {Mann}, {Mantelet}, {Marchal}, {Marchant}, {Marconi}, {Marie}, {Marinoni}, {Marrese}, {Marschalk{\'o}}, {Marshall}, {Mart{\'\i}n-Fleitas}, {Martino}, {Mary}, {Matijevi{\v{c}}}, {Mazeh}, {McMillan}, {Messina}, {Mestre}, {Michalik}, {Millar}, {Miranda}, {Molina}, {Molinaro}, {Molinaro}, {Moln{\'a}r}, {Moniez}, {Montegriffo}, {Monteiro}, {Mor}, {Mora}, {Morbidelli}, {Morel}, {Morgenthaler}, {Morley}, {Morris}, {Mulone}, {Muraveva}, {Musella}, {Narbonne}, {Nelemans}, {Nicastro}, {Noval}, {Ord{\'e}novic}, {Ordieres-Mer{\'e}}, {Osborne}, {Pagani}, {Pagano}, {Pailler}, {Palacin}, {Palaversa}, {Parsons}, {Paulsen}, {Pecoraro}, {Pedrosa}, {Pentik{\"a}inen}, {Pereira}, {Pichon}, {Piersimoni}, {Pineau}, {Plachy}, {Plum}, {Poujoulet}, {Pr{\v{s}}a}, {Pulone}, {Ragaini}, {Rago}, {Rambaux}, {Ramos-Lerate}, {Ranalli}, {Rauw}, {Read}, {Regibo}, {Renk}, {Reyl{\'e}}, {Ribeiro}, {Rimoldini}, {Ripepi}, {Riva}, {Rixon},
  {Roelens}, {Romero-G{\'o}mez}, {Rowell}, {Royer}, {Rudolph}, {Ruiz-Dern}, {Sadowski}, {Sagrist{\`a} Sell{\'e}s}, {Sahlmann}, {Salgado}, {Salguero}, {Sarasso}, {Savietto}, {Schnorhk}, {Schultheis}, {Sciacca}, {Segol}, {Segovia}, {Segransan}, {Serpell}, {Shih}, {Smareglia}, {Smart}, {Smith}, {Solano}, {Solitro}, {Sordo}, {Soria Nieto}, {Souchay}, {Spagna}, {Spoto}, {Stampa}, {Steele}, {Steidelm{\"u}ller}, {Stephenson}, {Stoev}, {Suess}, {S{\"u}veges}, {Surdej}, {Szabados}, {Szegedi-Elek}, {Tapiador}, {Taris}, {Tauran}, {Taylor}, {Teixeira}, {Terrett}, {Tingley}, {Trager}, {Turon}, {Ulla}, {Utrilla}, {Valentini}, {van Elteren}, {Van Hemelryck}, {van Leeuwen}, {Varadi}, {Vecchiato}, {Veljanoski}, {Via}, {Vicente}, {Vogt}, {Voss}, {Votruba}, {Voutsinas}, {Walmsley}, {Weiler}, {Weingrill}, {Werner}, {Wevers}, {Whitehead}, {Wyrzykowski}, {Yoldas}, {{\v{Z}}erjal}, {Zucker}, {Zurbach}, {Zwitter}, {Alecu}, {Allen}, {Allende Prieto}, {Amorim}, {Anglada-Escud{\'e}}, {Arsenijevic}, {Azaz}, {Balm}, {Beck}, {Bernstein},
  {Bigot}, {Bijaoui}, {Blasco}, {Bonfigli}, {Bono}, {Boudreault}, {Bressan}, {Brown}, {Brunet}, {Bunclark}, {Buonanno}, {Butkevich}, {Carret}, {Carrion}, {Chemin}, {Ch{\'e}reau}, {Corcione}, {Darmigny}, {de Boer}, {de Teodoro}, {de Zeeuw}, {Delle Luche}, {Domingues}, {Dubath}, {Fodor}, {Fr{\'e}zouls}, {Fries}, {Fustes}, {Fyfe}, {Gallardo}, {Gallegos}, {Gardiol}, {Gebran}, {Gomboc}, {G{\'o}mez}, {Grux}, {Gueguen}, {Heyrovsky}, {Hoar}, {Iannicola}, {Isasi Parache}, {Janotto}, {Joliet}, {Jonckheere}, {Keil}, {Kim}, {Klagyivik}, {Klar}, {Knude}, {Kochukhov}, {Kolka}, {Kos}, {Kutka}, {Lainey}, {LeBouquin}, {Liu}, {Loreggia}, {Makarov}, {Marseille}, {Martayan}, {Martinez-Rubi}, {Massart}, {Meynadier}, {Mignot}, {Munari}, {Nguyen}, {Nordlander}, {Ocvirk}, {O'Flaherty}, {Olias Sanz}, {Ortiz}, {Osorio}, {Oszkiewicz}, {Ouzounis}, {Palmer}, {Park}, {Pasquato}, {Peltzer}, {Peralta}, {P{\'e}turaud}, {Pieniluoma}, {Pigozzi}, {Poels}, {Prat}, {Prod'homme}, {Raison}, {Rebordao}, {Risquez}, {Rocca-Volmerange}, {Rosen},
  {Ruiz-Fuertes}, {Russo}, {Sembay}, {Serraller Vizcaino}, {Short}, {Siebert}, {Silva}, {Sinachopoulos}, {Slezak}, {Soffel}, {Sosnowska}, {Strai{\v{z}}ys}, {ter Linden}, {Terrell}, {Theil}, {Tiede}, {Troisi}, {Tsalmantza}, {Tur}, {Vaccari}, {Vachier}, {Valles}, {Van Hamme}, {Veltz}, {Virtanen}, {Wallut}, {Wichmann}, {Wilkinson}, {Ziaeepour}, \& {Zschocke}}]{Gaia2016}
{Gaia Collaboration}, {Prusti}, T., {de Bruijne}, J.~H.~J., {et~al.} 2016, \aap, 595, A1

\bibitem[{{Gaia Collaboration} {et~al.}(2023){Gaia Collaboration}, {Vallenari}, {Brown}, {Prusti}, {de Bruijne}, {Arenou}, {Babusiaux}, {Biermann}, {Creevey}, {Ducourant}, \& et~al.}]{Gaia2023}
{Gaia Collaboration}, {Vallenari}, A., {Brown}, A.~G.~A., {et~al.} 2023, \aap, 674, A1

\bibitem[{{Gehrels} {et~al.}(2004){Gehrels}, {Chincarini}, {Giommi}, {Mason}, {Nousek}, {Wells}, {White}, {Barthelmy}, {Burrows}, {Cominsky}, {Hurley}, {Marshall}, {M{\'e}sz{\'a}ros}, {Roming}, {Angelini}, {Barbier}, {Belloni}, {Campana}, {Caraveo}, {Chester}, {Citterio}, {Cline}, {Cropper}, {Cummings}, {Dean}, {Feigelson}, {Fenimore}, {Frail}, {Fruchter}, {Garmire}, {Gendreau}, {Ghisellini}, {Greiner}, {Hill}, {Hunsberger}, {Krimm}, {Kulkarni}, {Kumar}, {Lebrun}, {Lloyd-Ronning}, {Markwardt}, {Mattson}, {Mushotzky}, {Norris}, {Osborne}, {Paczynski}, {Palmer}, {Park}, {Parsons}, {Paul}, {Rees}, {Reynolds}, {Rhoads}, {Sasseen}, {Schaefer}, {Short}, {Smale}, {Smith}, {Stella}, {Tagliaferri}, {Takahashi}, {Tashiro}, {Townsley}, {Tueller}, {Turner}, {Vietri}, {Voges}, {Ward}, {Willingale}, {Zerbi}, \& {Zhang}}]{Gehrels2004}
{Gehrels}, N., {Chincarini}, G., {Giommi}, P., {et~al.} 2004, \apj, 611, 1005

\bibitem[{{Gianninas} {et~al.}(2010){Gianninas}, {Bergeron}, {Dupuis}, \& {Ruiz}}]{Gianninas2010}
{Gianninas}, A., {Bergeron}, P., {Dupuis}, J., \& {Ruiz}, M.~T. 2010, \apj, 720, 581

\bibitem[{{Hall} {et~al.}(2013){Hall}, {Tout}, {Izzard}, \& {Keller}}]{Hall2013}
{Hall}, P.~D., {Tout}, C.~A., {Izzard}, R.~G., \& {Keller}, D. 2013, \mnras, 435, 2048

\bibitem[{{Heber} {et~al.}(1984){Heber}, {Hunger}, {Jonas}, \& {Kudritzki}}]{Heber1984}
{Heber}, U., {Hunger}, K., {Jonas}, G., \& {Kudritzki}, R.~P. 1984, \aap, 130, 119

\bibitem[{{Hill} {et~al.}(2021){Hill}, {Lee}, {MacQueen}, {Kelz}, {Drory}, {Vattiat}, {Good}, {Ramsey}, {Kriel}, {Peterson}, {DePoy}, {Gebhardt}, {Marshall}, {Tuttle}, {Bauer}, {Chonis}, {Fabricius}, {Froning}, {H{\"a}user}, {Indahl}, {Jahn}, {Landriau}, {Leck}, {Montesano}, {Prochaska}, {Snigula}, {Zeimann}, {Bryant}, {Damm}, {Fowler}, {Janowiecki}, {Martin}, {Mrozinski}, {Odewahn}, {Rostopchin}, {Shetrone}, {Spencer}, {Mentuch Cooper}, {Armandroff}, {Bender}, {Dalton}, {Hopp}, {Komatsu}, {Nicklas}, {Ramsey}, {Roth}, {Schneider}, {Sneden}, \& {Steinmetz}}]{Hill2021}
{Hill}, G.~J., {Lee}, H., {MacQueen}, P.~J., {et~al.} 2021, \aj, 162, 298

\bibitem[{{Hippelein} \& {Weinberger}(1990)}]{Hippelein1990}
{Hippelein}, H., \& {Weinberger}, R. 1990, \aap, 232, 129

\bibitem[{{Horne}(1986)}]{Horne1986}
{Horne}, K. 1986, \pasp, 98, 609

\bibitem[{{Jones} {et~al.}(2022){Jones}, {Boffin}, {Brown}, {Zak}, {Hume}, {Munday}, \& {Miszalski}}]{JonesA702022}
{Jones}, D., {Boffin}, H. M.~J., {Brown}, A.~J., {et~al.} 2022, \mnras, 516, 4833

\bibitem[{{Jones} {et~al.}(2023){Jones}, {Hillwig}, \& {Reindl}}]{Jones2023}
{Jones}, D., {Hillwig}, T.~C., \& {Reindl}, N. 2023, in Highlights on Spanish Astrophysics XI, 216

\bibitem[{{Laseter} {et~al.}(2024){Laseter}, {Maseda}, {Curti}, {Maiolino}, {D'Eugenio}, {Cameron}, {Looser}, {Arribas}, {Baker}, {Bhatawdekar}, {Boyett}, {Bunker}, {Carniani}, {Charlot}, {Chevallard}, {Curtis-lake}, {Egami}, {Eisenstein}, {Hainline}, {Hausen}, {Ji}, {Kumari}, {Perna}, {Rawle}, {Rix}, {Robertson}, {Rodr{\'\i}guez Del Pino}, {Sandles}, {Scholtz}, {Smit}, {Tacchella}, {{\"U}bler}, {Williams}, {Willott}, \& {Witstok}}]{Laseter2024}
{Laseter}, I.~H., {Maseda}, M.~V., {Curti}, M., {et~al.} 2024, \aap, 681, A70

\bibitem[{{Liebert} {et~al.}(2013){Liebert}, {Bond}, {Dufour}, {Ciardullo}, {Meakes}, {Renzini}, \& {Gianninas}}]{Liebert2013}
{Liebert}, J., {Bond}, H.~E., {Dufour}, P., {et~al.} 2013, \apj, 769, 32, (L13)

\bibitem[{{Liebert} {et~al.}(1989){Liebert}, {Green}, {Bond}, {Holberg}, {Wesemael}, {Fleming}, \& {Kidder}}]{Liebert1989}
{Liebert}, J., {Green}, R., {Bond}, H.~E., {et~al.} 1989, \apj, 346, 251

\bibitem[{{Miller Bertolami}(2016)}]{millerbertolami16}
{Miller Bertolami}, M.~M. 2016, \aap, 588, A25

\bibitem[{{Miszalski} {et~al.}(2011){Miszalski}, {Acker}, {Parker}, {Boffin}, {Frew}, {Mikolajewska}, {Moffat}, \& {Napiwotzki}}]{Miszalski2011}
{Miszalski}, B., {Acker}, A., {Parker}, Q.~A., {et~al.} 2011, in Asymmetric Planetary Nebulae 5 Conference, P109

\bibitem[{{Miszalski} {et~al.}(2013){Miszalski}, {Miko{\l}ajewska}, \& {Udalski}}]{Miszalski2013}
{Miszalski}, B., {Miko{\l}ajewska}, J., \& {Udalski}, A. 2013, \mnras, 432, 3186

\bibitem[{{Osterbrock} \& {Ferland}(2006)}]{Osterbrock+06}
{Osterbrock}, D.~E., \& {Ferland}, G.~J. 2006, {Astrophysics of gaseous nebulae and active galactic nuclei}

\bibitem[{{Pengelly}(1964)}]{Pengelly1964}
{Pengelly}, R.~M. 1964, \mnras, 127, 145

\bibitem[{{Pereyra} {et~al.}(2013){Pereyra}, {Richer}, \& {L{\'o}pez}}]{Pereyra2013}
{Pereyra}, M., {Richer}, M.~G., \& {L{\'o}pez}, J.~A. 2013, \apj, 771, 114

\bibitem[{{Poole} {et~al.}(2008){Poole}, {Breeveld}, {Page}, {Landsman}, {Holland}, {Roming}, {Kuin}, {Brown}, {Gronwall}, {Hunsberger}, {Koch}, {Mason}, {Schady}, {vanden Berk}, {Blustin}, {Boyd}, {Broos}, {Carter}, {Chester}, {Cucchiara}, {Hancock}, {Huckle}, {Immler}, {Ivanushkina}, {Kennedy}, {Marshall}, {Morgan}, {Pandey}, {de Pasquale}, {Smith}, \& {Still}}]{Poole2008}
{Poole}, T.~S., {Breeveld}, A.~A., {Page}, M.~J., {et~al.} 2008, \mnras, 383, 627

\bibitem[{{Ramsey} {et~al.}(1998){Ramsey}, {Adams}, {Barnes}, {Booth}, {Cornell}, {Fowler}, {Gaffney}, {Glaspey}, {Good}, {Hill}, {Kelton}, {Krabbendam}, {Long}, {MacQueen}, {Ray}, {Ricklefs}, {Sage}, {Sebring}, {Spiesman}, \& {Steiner}}]{Ramsey1998}
{Ramsey}, L.~W., {Adams}, M.~T., {Barnes}, T.~G., {et~al.} 1998, in Society of Photo-Optical Instrumentation Engineers (SPIE) Conference Series, Vol. 3352, Advanced Technology Optical/IR Telescopes VI, ed. L.~M. {Stepp}, 34--42

\bibitem[{{Roming} {et~al.}(2005){Roming}, {Kennedy}, {Mason}, {Nousek}, {Ahr}, {Bingham}, {Broos}, {Carter}, {Hancock}, {Huckle}, {Hunsberger}, {Kawakami}, {Killough}, {Koch}, {McLelland}, {Smith}, {Smith}, {Soto}, {Boyd}, {Breeveld}, {Holland}, {Ivanushkina}, {Pryzby}, {Still}, \& {Stock}}]{Roming2005}
{Roming}, P. W.~A., {Kennedy}, T.~E., {Mason}, K.~O., {et~al.} 2005, \ssr, 120, 95

\bibitem[{{Schlafly} \& {Finkbeiner}(2011)}]{Schlafly2011}
{Schlafly}, E.~F., \& {Finkbeiner}, D.~P. 2011, \apj, 737, 103

\bibitem[{{Storey} \& {Hummer}(1995)}]{Storey-Hummer1995}
{Storey}, P.~J., \& {Hummer}, D.~G. 1995, \mnras, 272, 41

\bibitem[{{Tonry} {et~al.}(2012){Tonry}, {Stubbs}, {Lykke}, {Doherty}, {Shivvers}, {Burgett}, {Chambers}, {Hodapp}, {Kaiser}, {Kudritzki}, {Magnier}, {Morgan}, {Price}, \& {Wainscoat}}]{Tonry2012}
{Tonry}, J.~L., {Stubbs}, C.~W., {Lykke}, K.~R., {et~al.} 2012, \apj, 750, 99

\bibitem[{{Veras}(2021)}]{Veras2021}
{Veras}, D. 2021, in Oxford Research Encyclopedia of Planetary Science, 1

\bibitem[{{Werner} {et~al.}(2003){Werner}, {Deetjen}, {Dreizler}, {Nagel}, {Rauch}, \& {Schuh}}]{Werner2003}
{Werner}, K., {Deetjen}, J.~L., {Dreizler}, S., {et~al.} 2003, in Astronomical Society of the Pacific Conference Series, Vol. 288, Stellar Atmosphere Modeling, ed. I.~{Hubeny}, D.~{Mihalas}, \& K.~{Werner}, 31

\bibitem[{{Werner} {et~al.}(2018){Werner}, {Rauch}, \& {Kruk}}]{WernerEGB62018}
{Werner}, K., {Rauch}, T., \& {Kruk}, J.~W. 2018, \aap, 616, A73

\bibitem[{{Werner} {et~al.}(2024){Werner}, {Todt}, {Bond}, \& {Zeimann}}]{Werner2024}
{Werner}, K., {Todt}, H., {Bond}, H.~E., \& {Zeimann}, G.~R. 2024, arXiv e-prints, arXiv:2402.18976, (Paper III)

\bibitem[{{Willems} \& {Kolb}(2004)}]{Willems2004}
{Willems}, B., \& {Kolb}, U. 2004, \aap, 419, 1057

\bibitem[{{Wright} {et~al.}(2010){Wright}, {Eisenhardt}, {Mainzer}, {Ressler}, {Cutri}, {Jarrett}, {Kirkpatrick}, {Padgett}, {McMillan}, {Skrutskie}, {Stanford}, {Cohen}, {Walker}, {Mather}, {Leisawitz}, {Gautier}, {McLean}, {Benford}, {Lonsdale}, {Blain}, {Mendez}, {Irace}, {Duval}, {Liu}, {Royer}, {Heinrichsen}, {Howard}, {Shannon}, {Kendall}, {Walsh}, {Larsen}, {Cardon}, {Schick}, {Schwalm}, {Abid}, {Fabinsky}, {Naes}, \& {Tsai}}]{Wright2010}
{Wright}, E.~L., {Eisenhardt}, P. R.~M., {Mainzer}, A.~K., {et~al.} 2010, \aj, 140, 1868

\bibitem[{{Zuckerman} {et~al.}(1991){Zuckerman}, {Becklin}, \& {McLean}}]{Zuckerman1991}
{Zuckerman}, B., {Becklin}, E.~E., \& {McLean}, I.~S. 1991, in Astronomical Society of the Pacific Conference Series, Vol.~14, Astronomical Society of the Pacific Conference Series, ed. R.~{Elston}, 161--166

\end{thebibliography}

\end{document}